# Development and Validation of an Integrated LiDAR-Camera System for Real-Time Monitoring of Underground Longwall Operations

Pasindu Ranasinghe[1], Bikram Banerjee[2], Simit Raval[1]

[1]School of Minerals and Energy Resources Engineering, University of New South Wales, Sydney, NSW, Australia

[2]School of Surveying and Built Environment, University of Southern Queensland, Toowoomba, QLD, Australia

**Abstract**

Real-time spatial monitoring in underground longwall operations is challenging due to methane-related safety risks, poor visibility, elevated thermal loads, spatial confinement, and bandwidth-limited communications. Currently available camera-based monitoring provides visual context but lacks direct depth information, while standalone underground LiDAR scanners are limited to monochromatic or periodic 3D mapping. This paper presents the design, integration, and experimental validation of a LiDAR-camera monitoring system built around a certified flameproof enclosure that prevents flame propagation into the surrounding atmosphere. The system combines a solid-state LiDAR, an industrial RGB camera, and an onboard processor within a compact hardware assembly, supporting LiDAR-camera fusion, low-light image enhancement, and real-time processing. Laboratory experiments evaluated LiDAR and camera performance through the protective polycarbonate dome and quantified optical and geometric distortions introduced by the enclosure. Thermal testing showed that iterative component placement, heat sinking, and passive conduction reduced peak surface temperature from 106 °C to 70 °C, with internal temperature stabilising at 57 °C. Furthermore, a representative longwall simulation was created to evaluate the complete sensing, fusion, and transmission workflow under controlled geometric and low-light conditions. In the final configuration, more than 97% of LiDAR points fell within the camera field of view, supporting reliable colourisation. Enclosure-aware calibration and correction maintained geometric accuracy, while processed colourised point clouds were transmitted at up to 10 Hz with sustained bandwidth below 25 Mb/s. These results demonstrate the feasibility of a compact, thermally stable, and continuously operating LiDAR-camera monitoring system for real-time observation of underground longwall operations.

**Keywords:** LiDAR–camera fusion, Underground coal mining, Longwall monitoring, Real-time 3D monitoring, Flameproof enclosure, System integration

---

## 1. Introduction

Longwall coal mining is a highly productive method of underground coal extraction, widely adopted in modern mining operations. It involves a shearer cutting along an extended coal face supported by a series of movable hydraulic roof supports. Due to the scale and dynamic nature of these operations, workers were traditionally required to remain close to the coal face for production and inspection activities, exposing them to significant risks such as roof falls, heavy machinery, and airborne dust [1]. To mitigate these risks, the mining industry is increasingly shifting towards remote operation and automation [2]. This transition relies heavily on the availability of reliable, real-time monitoring systems that provide operators with accurate situational awareness without requiring physical presence at the mining face [3, 4].

However, developing such monitoring systems for underground coal mines is not straightforward. Methane and coal dust can create potentially explosive atmospheres, meaning that electronic equipment must either be intrinsically safe or housed within suitable explosion-protected enclosures [5]. These requirements restrict the type, size, power consumption, heat generation, cabling, and installation arrangement of sensors and onboard electronics. At the same time, underground longwall environments are affected by low illumination, airborne dust, humidity, vibration, restricted space, and moving machinery. These conditions reduce sensor visibility, complicate mechanical installation, increase the risk of signal degradation, limit available communication bandwidth, and make heat dissipation difficult within sealed housings. As a result, sensing systems used in longwall environments must not only capture useful monitoring data but also satisfy strict safety, integration, thermal, and communication constraints.

At present, underground monitoring systems largely rely on camera-based technologies due to their simplicity and widespread availability. Although 2D images, produced by these cameras, provide rich colour and texture information, they lack depth perception, which limits their use for accurate measurements and spatial analysis. As a result, essential spatial details such as precise measurements of distances, volumes, and the relative positions of objects cannot be directly obtained [3, 4]. Consequently, subtle yet critical features, including surface undulations, minor structural deformations, and small misalignments, can easily be overlooked.

Three-dimensional sensing technologies such as LiDAR (Light Detection and Ranging) can generate detailed maps of mine workings, allowing operators at the surface to visualise conditions as if they were standing next to the machinery underground [4]. This level of visibility helps in assessing ground conditions, verifying equipment alignment, and identifying hazards without requiring personnel to be physically present in high-risk areas. A notable example of an underground 3D-mapping sensor is CSIRO's ExScan device. It is the first and only 3D scanner certified for use in Australian coal mines [6]. While the ExScan has demonstrated the potential for LiDAR systems within mines, its output is limited to monochromatic point clouds and non-real-time visualisations. In practice, the ExScan requires manual triggering of scans, and its standalone sensor design results in reconstructions that suffer from gaps, voids, and a lack of colour information, leading to incomplete 3D representations.

Combining LiDAR with optical cameras can overcome these limitations by integrating geometric accuracy with visual context to produce real-time colourised 3D representations of the longwall face. However, in many existing systems, the sensors operate as independent or closed units, where LiDAR and camera data are acquired separately and must be transmitted to external systems for processing. Dedicated algorithms are then required to combine these data streams, rather than performing fusion within the sensing unit itself. This increases system complexity, communication bandwidth requirements, and processing latency, leading to delayed outputs that are unsuitable for dynamic and safety-critical underground mining environments.

Although advanced 3D sensing systems are increasingly used in construction, robotics, and surface mining, their adaptation to underground longwall monitoring requires more than selecting suitable sensors [4]. The sensing unit must combine LiDAR, imaging, onboard processing, enclosure-aware calibration, thermal control, and data transmission within a compact protected assembly. In longwall environments, this integration must also account for limited access for maintenance, sensor field-of-view alignment, heat buildup inside sealed housings, and the need to transmit processed data through existing underground communication networks. These requirements create a system-level integration challenge that is not fully addressed by conventional camera systems, standalone LiDAR scanners, or multi-sensor configurations that depend on external processing.

To address these challenges, this paper presents the design and experimental validation of a LiDAR–camera monitoring system for continuous, real-time observation of underground longwall operations. Unlike existing systems that rely on manual scan triggering or provide only monochromatic outputs, the proposed system delivers continuous, dense, colourised 3D representations of the longwall face to remote operators. The system is certified for safe operation in methane-hazardous environments and is housed within a flameproof enclosure. Implemented as a compact, self-contained unit mounted on the roof support at a fixed height and orientation (Figure 1), it performs onboard data fusion and transmission, reducing reliance on external processing infrastructure.

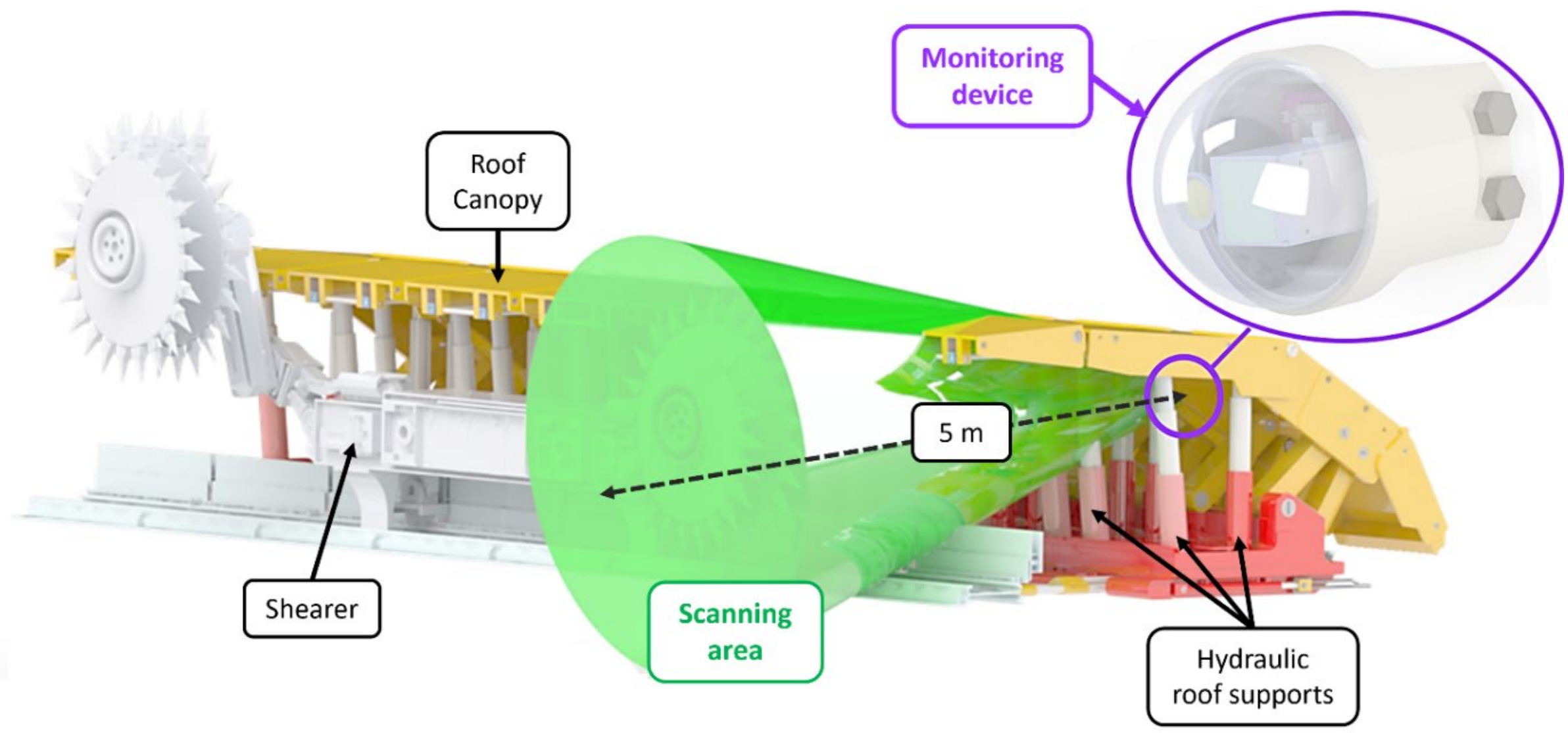

*Figure 1: Schematic of the proposed system mounted on the longwall roof support, showing the LiDAR scanning region (green cone) oriented toward the mining face.*

The main contributions of this work are summarised as follows:

1. A certified flameproof LiDAR–camera hardware platform for safe and continuous deployment in explosive underground coal mine environments, addressing mechanical, thermal, and regulatory constraints.
2. Experimental characterisation of enclosure-induced distortions, quantifying optical and geometric effects on LiDAR and camera measurements through a polycarbonate dome.
3. An optimised real-time onboard processing and fusion pipeline, incorporating temporal synchronisation, low-light enhancement, refraction correction, and colourised 3D reconstruction, while maintaining performance within available underground bandwidth and communication constraints.

4. Laboratory and simulation-based validation, demonstrating geometric accuracy, real-time performance, and reliable data transmission under representative longwall conditions.

The remainder of the paper is structured as follows: Section 2 outlines key design considerations for the system. Section 3 describes the system development, including hardware selection, system architecture, power assessment, iterative mechanical and thermal design, field-of-view alignment. Section 4 presents the sensor accuracy testing, including LiDAR and camera evaluation through the polycarbonate enclosure. Section 5 summarises the software pipeline for data acquisition, processing, and fusion. Section 6 presents the experimental validation results. Section 7 discusses the system performance, practical implications, future field validation, and potential analytical extensions. Section 8 concludes the paper with key findings.

## 2. Background

Designing an electronic monitoring system for an underground coal mine requires careful balancing of strict safety regulations and practical performance requirements. In this section, we discuss the key considerations that guided the development of the  hardware design:

1) Intrinsic Safety (IS) and Explosion Protection

All equipment used in underground coal mines must be certified for hazardous areas. This is critical because methane gas may infiltrate into electronic housings, and even a small spark or a high surface temperature can trigger a catastrophic explosion. As noted by industry experts, even seemingly harmless objects such as aluminium cans can pose ignition risks if crushed, highlighting the considerable challenges involved in deploying electronic equipment in such hazardous environments [7].

To comply with underground coal mine safety regulations, all components, including the LiDAR, camera, computer, and supporting electronics must be either intrinsically safe (IS) or housed in an explosion-proof enclosure [5]. Intrinsic safety means that the device operates at low energy levels, ensuring it cannot ignite methane-air mixtures, even under fault conditions. However, achieving IS certification for high-power electronics such as LiDAR and cameras is challenging as these devices often require more power and generate more heat than permitted under intrinsic safety standards. They also contain processors, lasers, capacitors, and inductors, which are difficult to isolate safely within the strict energy limits, making certification impractical.

The only practical and standards-compliant approach is to house the system within a flameproof enclosure under the Ex d protection concept as defined in IEC 60079-1, recognised under IECEx or ATEX certification schemes [8, 9]. These enclosures are designed to withstand and contain an internal ignition and to prevent the propagation of flame or hot gases to the surrounding explosive atmosphere. However, any external circuits, cables, glands, or data interfaces that penetrate or extend beyond the enclosure must themselves comply with intrinsic safety requirements to ensure that no ignition-capable energy is present outside the enclosure. These protection requirements, together with the enclosure's dimensions, mass, thermal limits, and cable-entry constraints, become critical factors in sensor selection, component placement, heat management, and the overall mechanical design of the system.

2) Sensor Requirements

Underground coal mines are extremely challenging for sensors. Airborne dust and humidity can attenuate laser beams and obscure optical images. To perform reliably in such conditions, LiDAR should operate at an infrared (IR) wavelength, which is less susceptible to scattering by dust particles. It must also have sufficient power and signal strength to penetrate moderate dust and achieve accurate detection at ranges of up to 5 metres. Having a multi-echo return capability can further enhance performance by allowing the detection of objects partially obscured by airborne particles [3, 10]. In parallel with LiDAR, the camera system must operate reliably under minimal illumination. This requires either integrated lighting or a high-sensitivity image sensor capable of low-light performance. The use of a large-aperture lens and pixel binning techniques can further improve the signal-to-noise ratio, enhancing image quality in dark underground conditions [11].

Beyond sensing capability, power consumption and heat dissipation are critical considerations, as the sensors are housed within a sealed, flameproof enclosure. High-power components increase internal temperatures, potentially leading to thermal shutdown, reduced reliability, or shortened component lifespan. Therefore, energy-efficient sensors with stable thermal characteristics are essential to ensure safe and continuous operation [12].

3) Power Requirements

All components must be carefully selected for energy efficiency to minimise heat generation within the sealed enclosure. Since active cooling methods that draw in fresh outdoor air are not possible, directly reducing power consumption helps maintain safe internal temperatures, improving system reliability and extending the operational life of the sensors.

Electrical power for the underground coal mines is supplied from the surface at high voltage and is stepped down through dedicated circuits and transformers to power the monitoring devices. The low-voltage supplies typically available are 120 VAC or 240 VAC outlets, together with 24 VDC and 12 VDC feeds. In longwall sections, both 24 VDC and 12 VDC may be distributed through certified intrinsically safe circuits in which voltage, current, and stored energy are limited in accordance with IEC 60079-11 [13]. However, the presence of an intrinsically safe supply does not by itself permit the connection of equipment. Any device connected to an IS circuit must be certified as compatible with the circuit parameters and protection concept. To comply with intrinsic safety requirements, certified IS barriers or galvanic isolators are used to limit voltage and current to safe levels and provide electrical isolation where required. These measures ensure that, under both normal and fault conditions, the available electrical and thermal energy remains below the ignition thresholds, thereby preventing ignition of methane-air or coal dust atmospheres [13, 14].

4) Range and Field of View

The monitoring system must effectively capture the longwall operation from an elevated position, focusing on the shearer, conveyor system, and roof supports. Previous longwall roadway studies have reported roadway cross-sectional dimensions of approximately 3.5–4.5 m, highlighting the confined spatial scale in which underground monitoring systems must operate. Accordingly, a 5 m monitoring envelope was adopted as a conservative short-range design basis for selecting the LiDAR range and scanning pattern [15]. Within this monitoring envelope, a wide-field-of-view solid-state LiDAR is a suitable choice over a 360° spinning unit, as it allows for focused scanning of the longwall face and surrounding structures while avoiding unnecessary data from non-critical directions, such as machinery housing or the goaf area behind the sensor. Capturing such redundant data adds significant processing load without contributing to the primary front-facing monitoring objectives, thereby making directional LiDAR a more efficient choice. To generate a complete and detailed representation, the camera's lens FOV must substantially overlap with the LiDAR's coverage. This alignment provides complementary colour and texture information that enhances the interpretability of the LiDAR's 3D geometry.

5) Real-time Continuous Monitoring

The monitoring system should continuously capture, process, and transmit data with minimal latency to maintain real-time awareness, as underground conditions can change rapidly and hazards may develop within seconds. However, acceptable latency in underground mining is strongly application-dependent [16]. For example, some IoT-based safety monitoring systems are designed to operate with communication delays below approximately 20 ms to support rapid hazard detection and response [17]. In contrast, broader teleoperation studies indicate that higher latencies may still be acceptable depending on the task, with remote operation scenarios often functioning effectively at around 100 ms, and in some controlled or slower-paced tasks, delays of up to 0.4–0.5 s remain usable [18].

6) Data Transmission

Underground mines combine multiple communication backbones to move data from the longwall face to a remote operations centre. These include fibre-optic Ethernet run along the longwall, power-line Ethernet carried on trailing cables, leaky-feeder coax systems for VHF/UHF radios, Wi-Fi access points installed in the gate roads, and, more recently, emerging 5G networks [19-21]. Any monitoring device intended for these environments must comply with and seamlessly integrate into the existing mine communication infrastructure to ensure efficient data flow without requiring major modifications.

However, available bandwidth at the longwall face can be constrained, particularly where legacy leaky-feeder or wireless systems are used. Continuous transmission of raw high-resolution image streams and dense LiDAR point clouds may generate substantial data rates. For example, compressed 4K video streams can require 15–50 Mb/s depending on encoding settings, while high-density LiDAR data streams may add several additional megabits per second [22]. Therefore, on-board data processing, compression, and optimisation are essential to reduce transmission load and ensure compatibility with typical underground network capacities.

7) Integration and Installation

Longwall operations are dynamic and space-constrained, requiring monitoring systems that are compact, easily mountable, and minimally intrusive to existing infrastructure. The system should be designed for secure mounting on roof supports or adjacent structures without interfering with operational equipment. Minimising external cabling and auxiliary components simplifies installation and reduces potential failure points in environments subject to vibration, impact, and mechanical movement. Furthermore, the assembly must allow straightforward integration, removal, and repositioning to facilitate routine maintenance, system upgrades, and relocation as the longwall face advances.

8) Scene Representation Requirements

A detailed representation of the mine environment is essential for effective monitoring and analysis. This representation must capture precise measurements of distances, geometries, and structural features—including roof mesh, support members, and equipment—to support comprehensive integrity assessments and timely hazard identification. Additionally, the system should incorporate full-colour information into the 3D point cloud to differentiate materials and reveal subtle anomalies that would not be apparent in a monochromatic data stream [23].

In summary, these considerations shaped a design philosophy to create a rugged, safe, integrated, and efficient monitoring system. Next, we discuss how specific hardware components were selected in line with these requirements.

## 3. The System Development

### 3.1 Hardware Selection

Based on the above considerations, we carefully selected the main hardware components of the system: the LiDAR sensor, the camera, and the processing computer, along with the supporting elements for integration. The following sections describe the chosen components and their key specifications, with Figure 2 illustrating the system architecture with all components and connections.

**LiDAR:** We required a LiDAR capable of providing dense 3D point clouds of the longwall environment while fitting within an explosion-proof housing. The Livox Avia LiDAR sensor was selected after an extensive review of available units. This compact solid-state LiDAR provides high point-throughput and offers a maximum detection range of up to 190 m for 10% reflectivity targets (and around 320 m for high-reflectivity surfaces), exceeding the operational requirements. Its FOV is approximately 70° horizontal × 77° vertical in its standard non-repetitive scanning mode. This forward-facing coverage aligns with the planar and linear geometry of the longwall mining face, where full 360° scanning is unnecessary.

The sensor's triple-echo return capability allows it to detect multiple reflections from a single laser pulse, which helps in identifying objects that are partially obscured by airborne particles. The Livox Avia can output up to 720,000 points per second, producing dense point clouds. Its non-repetitive, pseudo-random scanning pattern gradually fills a circular region over time, enabling a more uniform point distribution across the entire FOV. This is advantageous for capturing fine details, as stationary scanning gaps are filled by accumulating multiple frames. Additionally, the Livox Avia includes an integrated IMU (Inertial Measurement Unit), which we used to monitor the orientation of the entire system.

The Livox Horizon—a similar solid-state unit and the predecessor to the Livox Avia—was found to be highly effective for forward-looking scans in underground mines [10]. The same study showed that solid-state lidars introduce fewer blind spots compared to mechanical lidars, which often require multiple units for complete coverage due to structural obstructions. They also demonstrated that solid-state LiDARs could operate reliably in dusty, muddy, and waterlogged environments [10], further supporting the suitability of the selected sensor for this application.

**Camera:** An industrial-grade digital camera, the FLIR Blackfly S, was chosen as the vision sensor. It features a 2.3 MP CMOS global-shutter sensor, which is important for avoiding motion blur or distortion in images when the device itself moves or when there are moving parts in the scene. The FLIR camera streams at up to 43 frames per second at 1936 x 1464 pixel resolution. It has a pixel size of 4.5 x 4.5 µm. This larger pixel size allows each pixel to capture more light, making it highly suitable for low-light imaging. The sensor offers a dynamic range of 70.63 dB, enabling it to extract detail from both dark and bright regions in the scene. We applied pixel binning (combining pixels) at a factor of 4, effectively improving visibility in near-darkness, though at the cost of some image resolution.

The camera, paired with an Edmund Optics wide-angle 4 mm f/1.8 lens, offers a wide field of view (~101.6° diagonally) that exceeds the LiDAR's coverage. Additionally, we developed a software-based low-light enhancement algorithm to further improve image clarity without requiring active lighting [24].

**Processing Computer:** The heart of the system is the onboard computer that collects sensor data, runs processing algorithms, and communicates with the external network. We selected a Single Board Computer (SBC) – specifically the Radxa Rock 5 Model A – as the processing unit. The SBC is based on the Rockchip RK3588 system-on-chip (SoC), featuring an 8-core 64-bit ARM CPU, an integrated neural processing unit (NPU), and a GPU, supported by 8 GB of RAM. The choice was guided by the need for a compact yet powerful computing platform that could fit inside the enclosure. The Rock 5 has low power consumption relative to full-sized PCs, helping with thermal management, yet it is powerful enough to run our Robot Operating System (ROS) based software and handle point cloud and image data in real time. We considered other options such as the NVIDIA Jetson series, LattePanda computers, and Raspberry Pi 5, but the Rock 5 provided a good balance of CPU performance while fitting within the required boundaries of the enclosure.

In our configuration, the SBC runs the Ubuntu 20.04 LTS operating system with ROS middleware to handle sensor communication. It receives GigE data streams from both the camera and the LiDAR. However, the SBC had only a single native Gigabit Ethernet port, whereas both sensors required dedicated high-throughput Ethernet connections. Our first attempt involved using a passive network splitter to share the single Ethernet port. This approach failed, as the splitter could not manage bidirectional data streams independently. Without proper traffic isolation, the system experienced frequent data collisions and packet loss, making it unusable for synchronised sensor acquisition. We then explored a USB3-to-Ethernet adapter to create a second network interface. While this offered physical separation, we encountered significant driver compatibility issues and inconsistent performance due to USB bandwidth constraints. To overcome these limitations, we integrated a compact 4-port industrial Ethernet switch within the enclosure. This allowed both the LiDAR and the camera to connect simultaneously to the switch, which then routed their combined data through a single uplink to the SBC's Ethernet port. This configuration proved effective – it allowed simultaneous high-throughput reception from the LiDAR and camera without dropping packets and provided an additional port to be used as an interface for sending data to the outside.

For data storage and logging, the SBC includes an onboard eMMC drive, allowing raw sensor data to be recorded locally during experiments or incident investigations, even when the connection to the monitoring station is temporarily unavailable.

### 3.2 The System Architecture

All components were assembled, and the connection architecture is shown in Figure 2. The device is powered by the mine's intrinsically safe 12 VDC line, which first passes through an intrinsically safe power barrier within the enclosure. This barrier limits current and energy to prevent ignition and ensures electrical isolation. The regulated power is then distributed to the LiDAR, camera, onboard computer, and the network switch. Both sensors connect via Ethernet to the onboard SBC through an internal network switch. The SBC processes the incoming data streams in real time, fusing LiDAR point clouds with synchronised camera images to generate dense, colourised 3D outputs. These processed data are then transmitted through an IS data barrier to ensure compliance with underground safety regulations before reaching the surface monitoring station for visualisation and analysis.

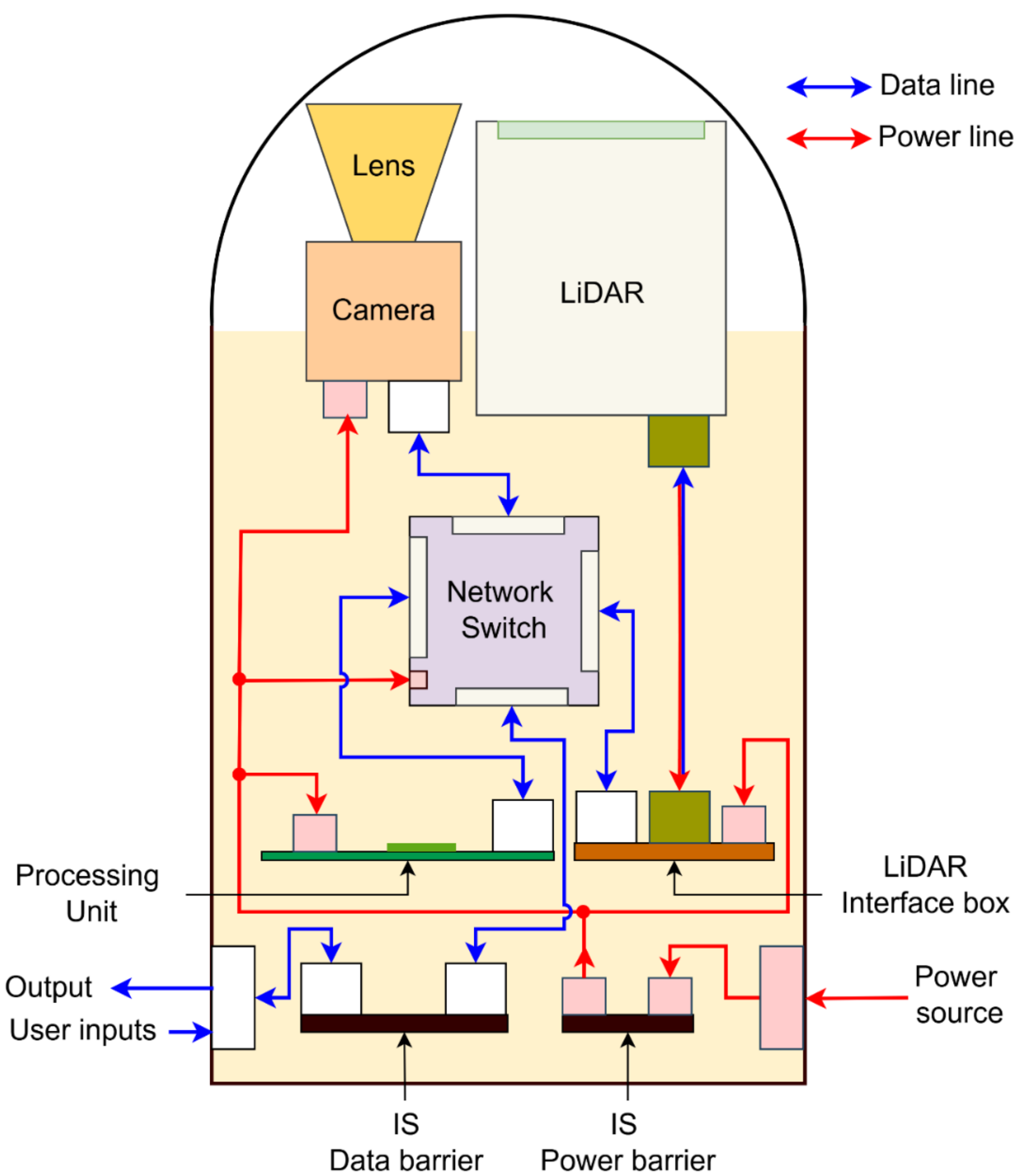

*Figure 2: Hardware architecture of the proposed monitoring system, mounted inside the enclosure. The LiDAR and camera feeds are routed to the integrated processing unit via an internal Ethernet switch. The system consists of a LiDAR, camera, processing unit, and network switch, all interfaced through an IS-compliant power-and-data barrier to ensure safe operation in underground mining conditions.*

All components are housed within a certified flameproof cylindrical enclosure, manufactured from high-strength steel to withstand internal explosion pressures in accordance with IEC 60079-1 requirements. It features a hemispherical optical-grade polycarbonate dome at the sensing end, providing a wide and unobstructed viewing envelope for both the LiDAR and camera sensors. The internal component layout was designed using custom 3D CAD models and refined through iterative simulations, verifying that all components fit within the confined space while maintaining the necessary field of view, ease of maintenance, thermal management, and overall system stability. Two sealed connections—one for power and one for Ethernet data—are the only interfaces with the external environment and are designed to maintain explosion protection and system integrity in hazardous underground conditions.

**Power Requirement:** The selected enclosure is rated for a maximum internal power consumption of 40 W. To ensure compliance, the total power requirement of the system was carefully evaluated. The results are summarised in Table 1. During start-up, the total power consumption can reach 34.2 W, but it stabilises below 25 W under operation, keeping the system within safe operating limits.

*Table 1: Power and startup characteristics of the core system components.*

| Component | Rated Input Voltage Range (V) | Input Voltage (V) | Bootup time (s) | Power (W) Start-up | Power (W) – Max Operational |
|---|---|---|---|---|---|
| Livox Avia LiDAR | 9 - 30 | 12 | 12 | 14 | 7.8-8 |
| FLIR Blackfly S camera | 12-20 | 12 | 12-15 | 4.2 | 2.8 |
| Rock 5 Model A SBC | 5-20 | 12 | 20 | 15 | 13 |
| Network Switch | 12 | 12 | - | <1 | <1 |
| | | | | 34.2 | 24.8 |

### 3.3 Mechanical Integration and Thermal Optimisation

The development process involved multiple cycles of mechanical integration, thermal evaluation, and design refinement. The main objective was to package the LiDAR, camera, single-board computer, network switch, LiDAR interface module, power barrier, and data barrier within the limited internal volume of the cylindrical flameproof enclosure. Particular emphasis was placed on maintaining an unobstructed sensor view through the polycarbonate dome, reducing heat accumulation within the sealed enclosure, simplifying electrical routing, and improving maintenance access.

A detailed 3D model of the system's internal assembly was developed using SolidWorks software. The initial objective was to arrange the largest and most space-constrained components—LiDAR, camera, single-board computer (SBC), network switch, and the LiDAR interface module (a custom board that manages power and data connectivity for the LiDAR)—in a compact, efficient configuration. Figure 3 shows the resulting layout, where these major components, along with all other internal elements, were stacked and mounted on custom brackets within the cylindrical flameproof enclosure.

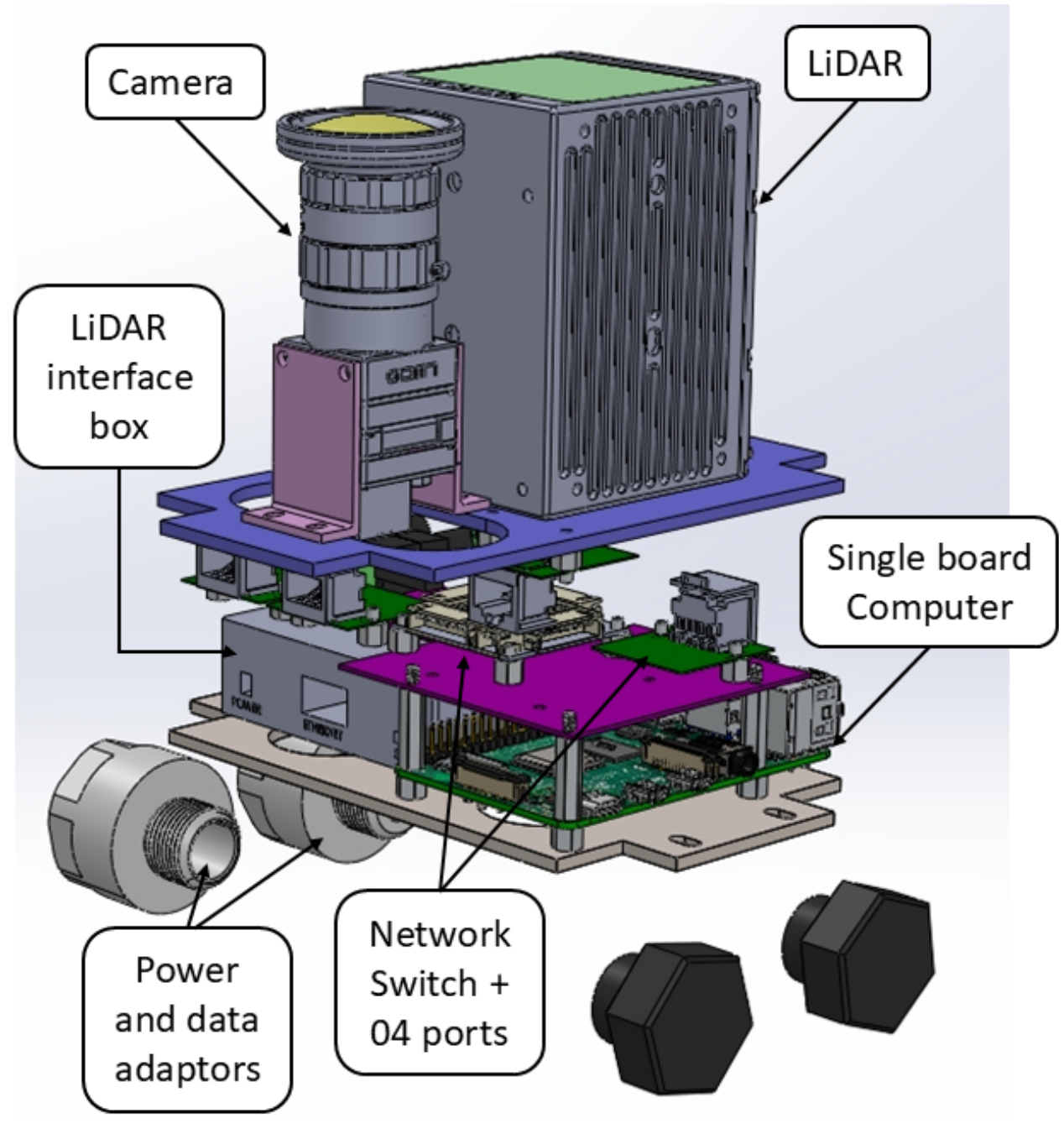

*Figure 3: Initial hardware assembly*

The LiDAR and camera were mounted on the top bracket, aligning them to look outward through the transparent polycarbonate dome. This bracket was secured in a dedicated groove within the enclosure. In the lower section, the SBC and LiDAR interface box were mounted on a separate plate, which was fastened to existing holes in the enclosure. A secondary platform, supported by standoffs from the SBC mount, was used to hold the network switch. Separate network ports provided dedicated interfaces for each module. All supporting brackets and plates were fabricated from aluminium due to its lightweight and ease of machining. Although aluminium is typically avoided in underground coal mines due to its potential to generate sparks, its use is permissible within a sealed flameproof enclosure, where it remains isolated from the mine atmosphere and poses no ignition risk [25].

Once the system was fully assembled, initial power-on tests and functional checks were conducted, followed by an evaluation of its thermal behaviour due to expected heat buildup within the sealed enclosure. A temperature sensor was installed at the rear of the lower mounting plate to capture overall internal heat accumulation, while a FLIR E6 thermal camera was used to visualise surface temperature distributions both inside and outside the enclosure, enabling identification of localised hot spots. As shown in Figure 4, the system was placed inside a controlled-temperature oven maintained at 40 °C to replicate worst-case underground mine conditions, and temperature readings were recorded at 3-hour intervals over a 15-hour period.

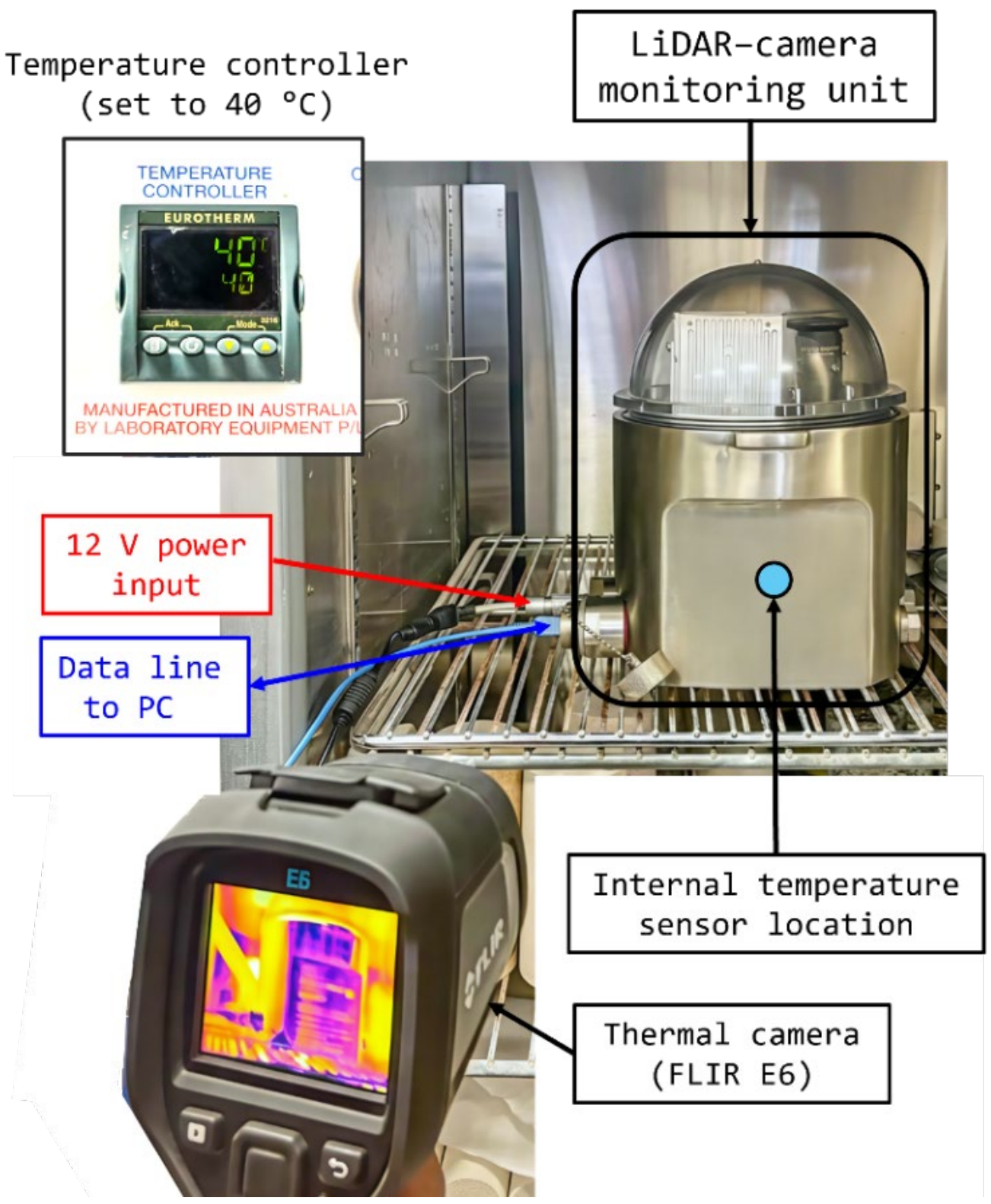

*Figure 4: Experimental setup for thermal evaluation of the LiDAR–camera monitoring system. The monitoring unit is placed inside a controlled-temperature oven maintained at 40 °C, with 12 V power and data connections active. A FLIR E6 thermal camera is used to capture surface temperature distributions during testing.*

Thermal imaging showed a steady rise in temperature, with the camera surface reaching a peak of 106°C, while the internal temperature sensor recorded a maximum of 77.6°C. Despite exceeding the safe limits for many electronic components, the system continued to operate reliably throughout the test. No failures or performance degradation were observed; notably, no CPU throttling occurred, indicating stable processing performance even under elevated thermal conditions.

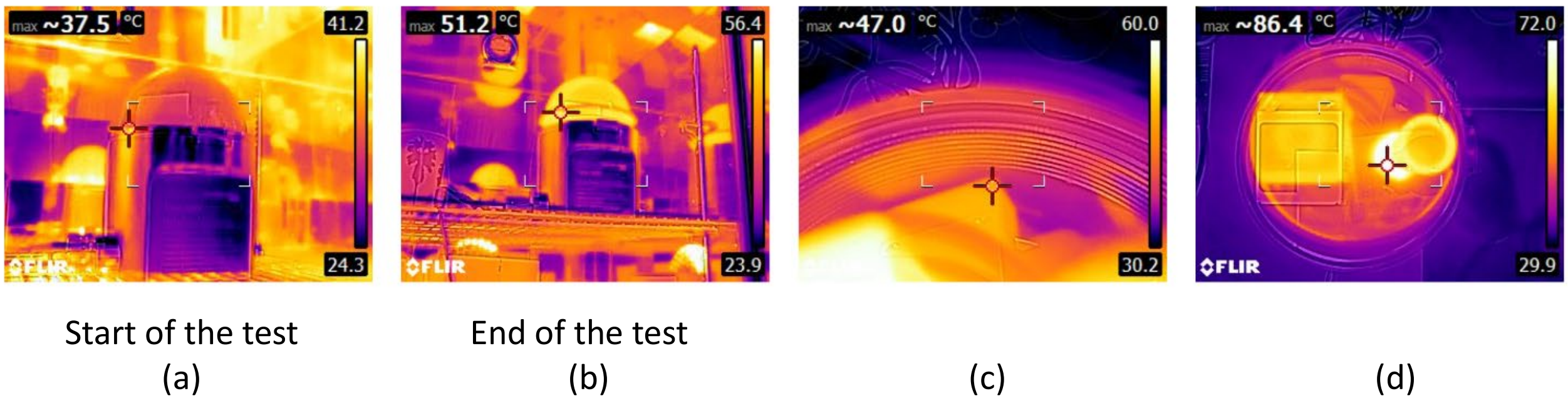

*Figure 5: Thermal evaluation of the initial design (a) Temperature profile at the beginning of the experiment (b) External temperature profile after 15 hours, showing that the transparent dome is warmer than the base (c) Internal wall temperature profile after 15 hours (d) Top-down thermal image highlighting the camera as the primary heat source.*

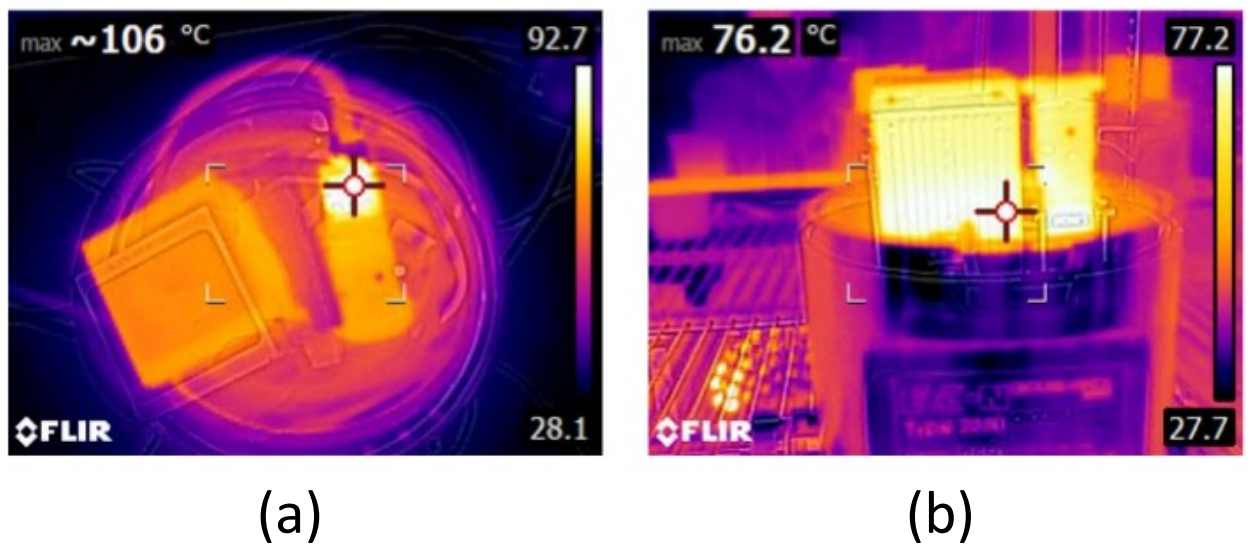

(a) (b)

*Figure 6: Maximum temperatures recorded for (a) the Camera (b) the LiDAR during the initial thermal tests.*

Analysis of the thermal images identified the camera as the primary heat source, followed by the LiDAR and the SBC. The highest surface temperatures were observed around the camera's network port region, as shown in Figure 5(d) and Figure 6(a). To mitigate this, the camera was powered via a separate 12 V DC supply (Figure 2), replacing the previous Power over Ethernet (PoE) configuration. Heat sinks were attached to two sides of the camera, while thermal pads were applied to the remaining surfaces to establish thermal contact with the mounting bracket. As the bracket is directly coupled to the enclosure wall, this effectively enabled the enclosure to act as a passive heat sink. Similar heat dissipation strategies were applied to the LiDAR to improve thermal conduction.

In addition, the internal layout was reconfigured to increase air volume around heat-generating components. The SBC was repositioned away from the camera and LiDAR to reduce thermal coupling. A heat sink was added to the SBC CPU, and a small cooling fan was mounted above it to improve airflow and reduce localised heat buildup.

Iterative rounds of thermal testing were conducted under identical conditions. In the second design iteration, the maximum steady-state surface temperature obtained from the thermal images was reduced to 86 °C, and further decreased to 70 °C in the final design, as shown in Figure 7, while the internal temperature measured by the sensor dropped to 57 °C. Although this remains relatively high, all components are rated for operation above this level, placing the system within acceptable limits. The temperature rise on the metal enclosure walls was minimal; however, the polycarbonate dome reached a maximum of approximately 51 °C, similar to the initial design. Following the implemented modifications, the thermal profiles of the camera and LiDAR became closely aligned, indicating a more balanced heat distribution, with the LiDAR exhibiting a slightly higher peak surface temperature of 70.3 °C compared to 69.8 °C for the camera, as shown in Figure 7(b) and Figure 7(c).

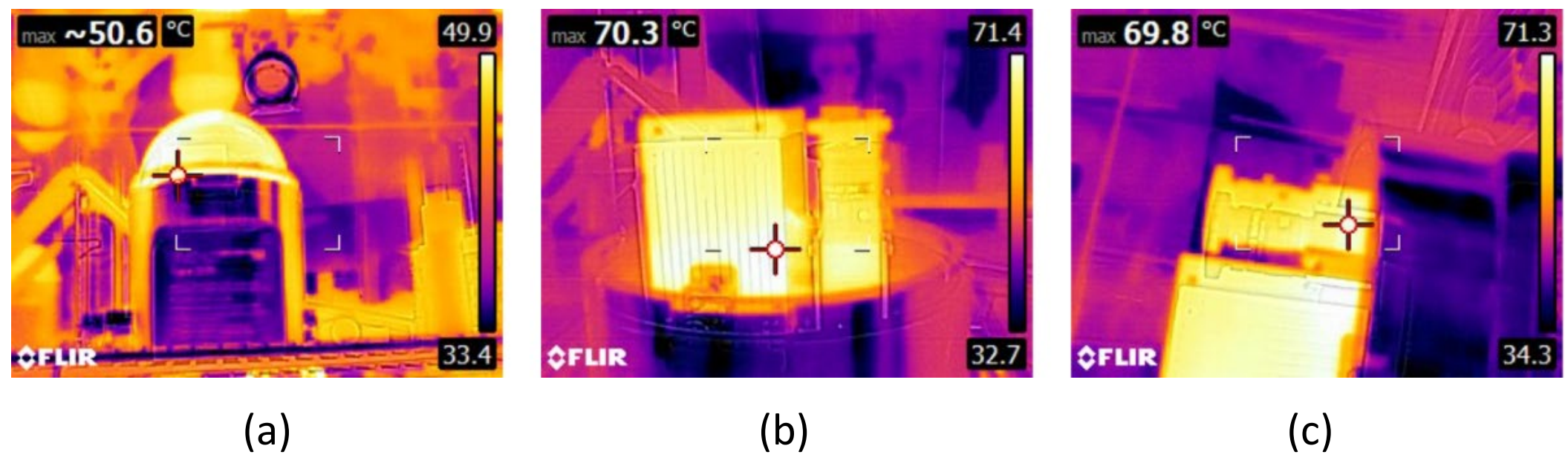

(a) (b) (c)

*Figure 7: Thermal evaluation of the final design temperature profiles after 15 hours (a) External Temperature, showing that the transparent dome is warmer than the base (b) Internal component temperature profile indicating that the LiDAR is now the primary heat source (c) Highest temperature regions observed on the camera surface.*

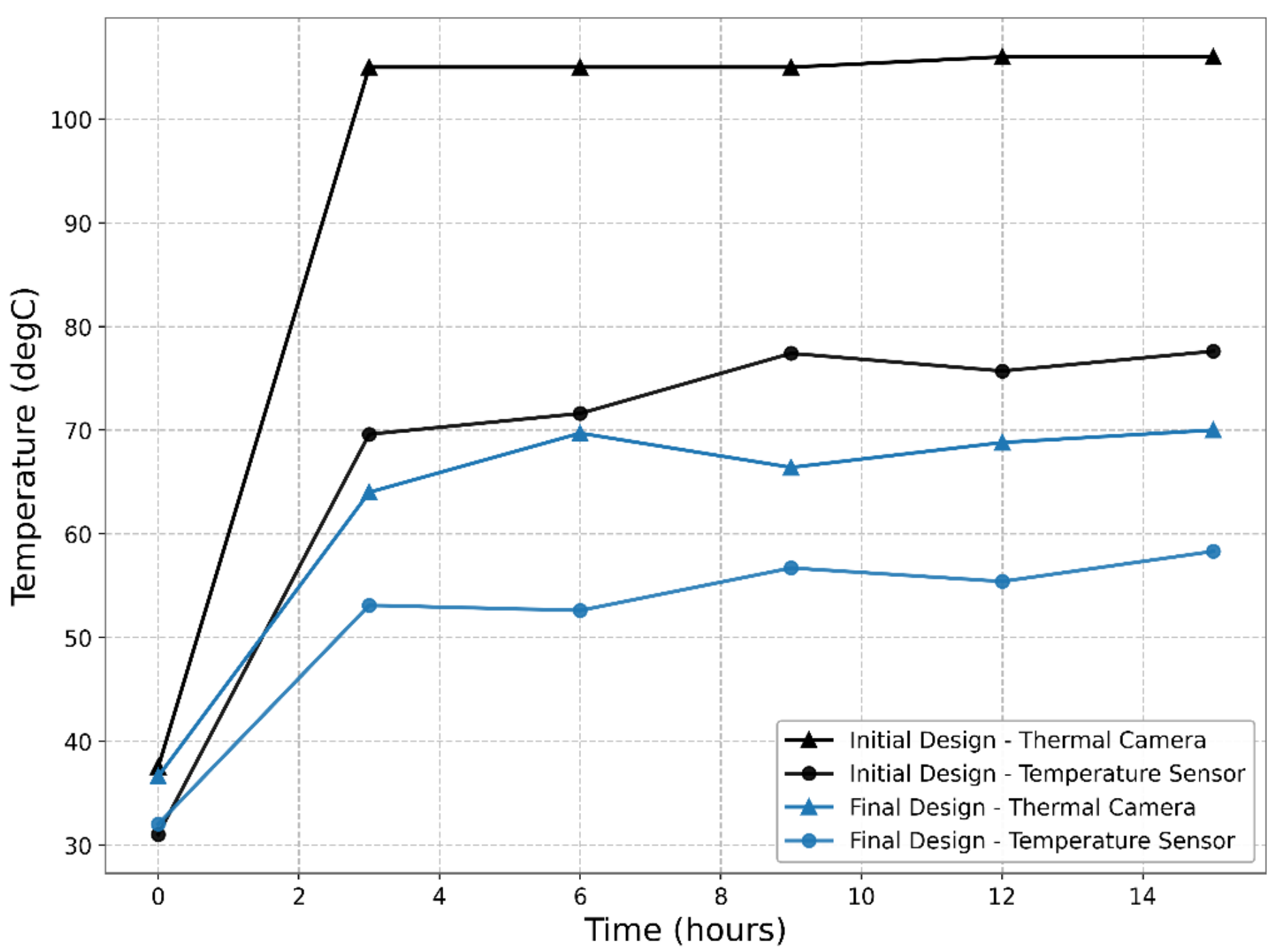

*Figure 8: Temperature profiles over time for the initial design (black lines) and final design (blue lines). Both camera (triangle markers) and internal temperature sensor (circle markers) readings are shown.*

As shown in Figure 8, the system reaches its steady-state temperature within the first three hours and maintains thermal equilibrium throughout operation. The final design shows a clear improvement in thermal performance, with a reduction of approximately 35 °C at the surface and 20 °C internally, demonstrating the effectiveness of the iterative design modifications. The final configuration maintains a stable thermal profile and operates within the safe limits of all internal components. Furthermore, a continuous CPU temperature monitoring mechanism with an automatic shutdown threshold of 80 °C was implemented, providing an additional layer of protection.

Following the thermal optimisation, the internal hardware layout was finalised to support sensor visibility, heat dissipation, electrical routing, and maintainability. The camera and LiDAR occupy the top layer to maintain an unobstructed field of view. The SBC and LiDAR interface box are placed on a second layer. Intrinsically safe power and data barriers are housed in the lower section, away from the main heat sources. Internal wiring has been replaced with a custom PCB, reducing cable clutter and lowering the risk of connection failures. This PCB, mounted above the SBC, efficiently routes both power and data lines, creating a cleaner system.

A one-piece mounting bracket design was adopted, bringing all supporting brackets together on a single larger contact surface within the enclosure. This change improves both mechanical integration and thermal performance by avoiding reliance on multiple small contact areas. Additionally, maintenance and hardware upgrades become more straightforward since the entire assembly can be removed or reattached as a single unit, preserving the calibrated alignment of the LiDAR and camera relative to the enclosure.

**3.4 Field of View Alignment and Coverage Analysis**

After addressing thermal concerns, we proceeded to validate both the individual and combined fields of view of the sensors to ensure effective sensor fusion with minimal blind spots. Although the camera is equipped with a 4 mm lens, the use of a 2/3″ sensor results in an effective field of view equivalent to that provided by a 16 mm lens on a full-frame camera. This configuration provides an effective horizontal FOV of ~ 84.8°, a vertical FOV of ~ 65.3°, and a diagonal FOV of ~ 101.6°. These values define a viewing frustum that, when projected onto a plane, forms a rectangular region. In contrast, the Livox Avia LiDAR produces a conical scanning volume with a horizontal FOV of 70° and a vertical FOV of 77°. This conical field of view appears as an elliptical cross-section. To maximise overlap and ensure that all LiDAR points can be colourised, the system is oriented so that the LiDAR's horizontal axis aligns with the camera's vertical axis, while the camera's wider horizontal axis fully covers the LiDAR's vertical span. This alignment significantly improves the spatial overlap between the two sensors.

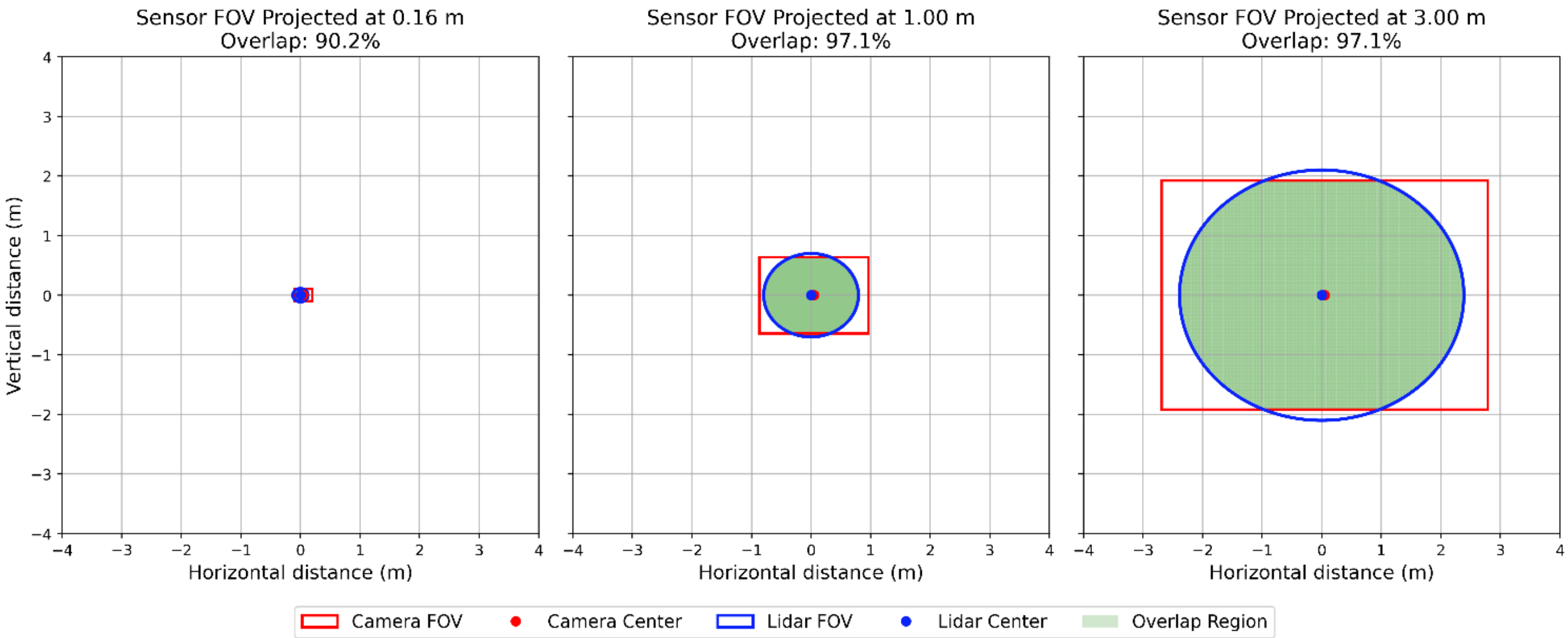

*Figure 9: FOV analysis: Camera (red rectangle) and LiDAR (blue ellipse) fields of view at three different distances (0.16 m, 1.0 m, and 3.0 m). The sensor centres are offset by 5 cm horizontally, with no vertical offset. The green area highlights their overlap region.*

As shown in Figure 9, at a distance of 16 cm, approximately 90% of the LiDAR points already fall within the camera's FOV. By 25 cm, 95% of LiDAR points are within the camera's view. Beyond the 30 cm point, the system consistently maintains 97% overlap. Although a small region at the top and bottom of the LiDAR's scanning area is not covered, and a slight blind spot exists between the two sensors due to their horizontal offset, these areas are negligible. Notably, if the camera and LiDAR had been aligned conventionally, with matching horizontal and vertical axes, the maximum overlap would be 90%, and it would require a distance of 2.05 m to achieve that. Thus, this careful alignment strategy increases the effective scanning area by ~7% and significantly enhances the system's working range.

## 4. Sensor Accuracy Testing

### 4.1 LiDAR

Once the system's field of view was validated, the performance of each sensor was evaluated under realistic operating conditions, specifically with observations made through the dome-shaped polycarbonate enclosure. A controlled experiment was conducted using a rectangular target of known dimensions, positioned at fixed distances of 2 m, 3 m, 4 m, and 5 m from the LiDAR. At each distance, point cloud data were collected at 10 Hz for 2 s, resulting in 20 frames per test. The height, width, and distance to the target were derived from the resulting point clouds under two conditions: with and without the transparent enclosure in place. The measurements obtained with the enclosure were then compared with the corresponding reference measurements obtained without the enclosure, providing a basis to assess how the dome influences dimensional and range accuracy. The relative measurement error for each measurement was calculated using the following expression:

$$\text{Relative measurement error (\%)} = \left(\frac{\mathrm{M_e} - \mathrm{M_r}}{\mathrm{M_r}}\right) \times 100\%$$

where $M_e$ represents the measurement obtained with the enclosure, and $M_r$ represents the reference measurement obtained without the enclosure. The results of this analysis are presented in Figure 10.

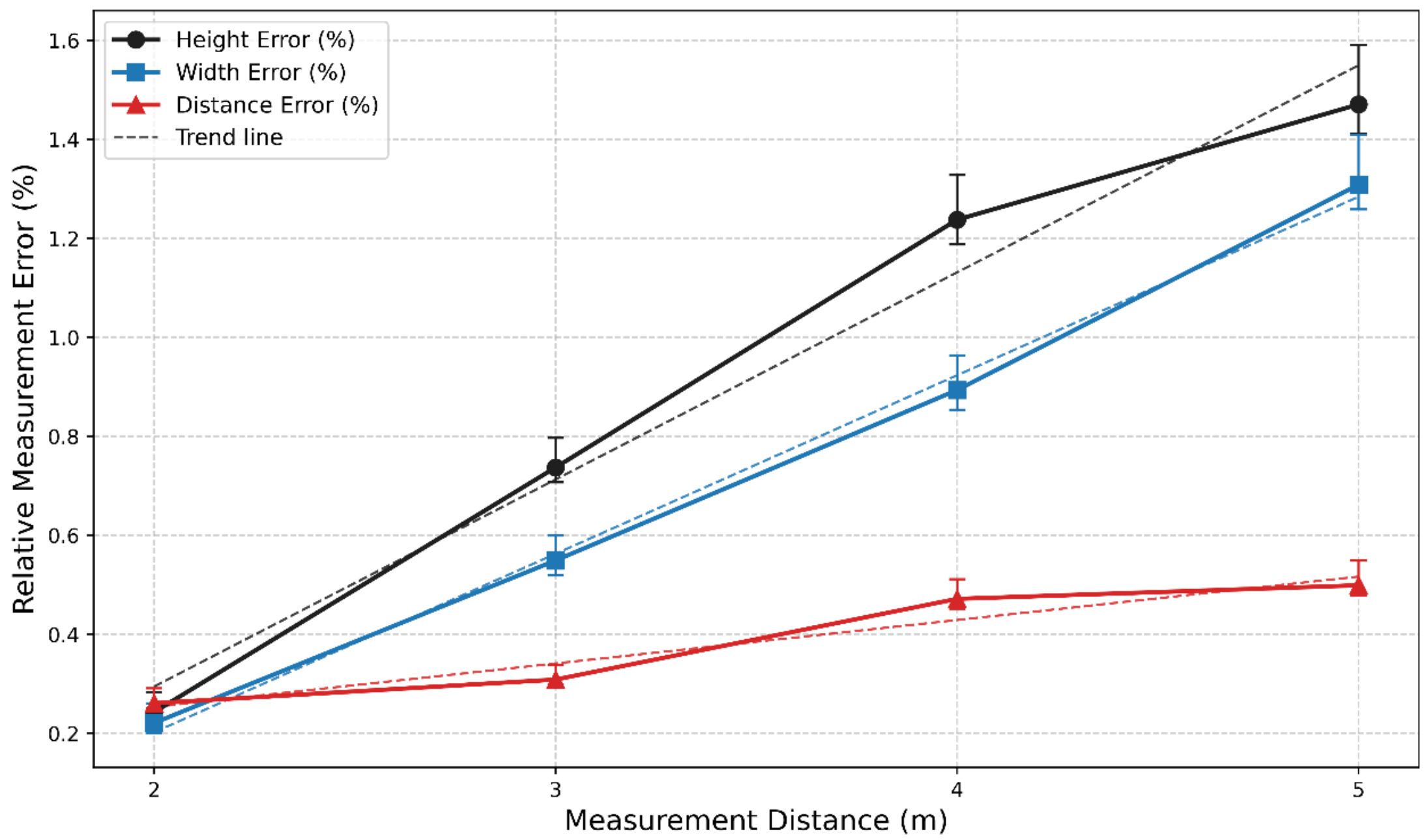

*Figure 10: Relative measurement errors in object height, width, and distance—estimated from point clouds with the enclosure—compared across varying LiDAR-to-object distances.*

The results, as illustrated in Figure 10, indicate a consistent positive deviation across all measured parameters when the enclosure is present, suggesting that the dome introduces systematic overestimation in the reconstructed dimensions. The error bars, representing one standard deviation across repeated frames, remain small across all distances, indicating stable measurements under the tested conditions. However, the variability increases slightly with range, with the largest error bars observed at 5 m, particularly for height and width estimation. Among the evaluated quantities, height error shows the most pronounced increase with measurement distance. This trend is mainly attributed to the larger vertical divergence of the LiDAR beam, 0.28°, which causes progressive spreading of the laser footprint as range increases.

In contrast, width error increases more gradually, reflecting the smaller horizontal beam divergence of 0.03°. The reduced lateral spreading limits distortion in this direction, making width measurements less sensitive to enclosure-induced effects. However, the variation in error does not follow a strictly linear relationship with the magnitude of beam divergence, suggesting that additional factors such as dome curvature, incidence angle, and point distribution on the target surface also contribute to the observed distortion. Range measurements show a different behaviour, remaining relatively stable across all evaluated distances. This is expected because LiDAR range estimation is governed primarily by time-of-flight principles, which are less affected by the enclosure than angular beam deviation [26]. The dome mainly introduces a small and nearly constant optical path offset through the polycarbonate material, resulting in a uniform bias and the near-flat trend observed in the relative distance error.

These results represent the baseline geometric distortion introduced by the polycarbonate enclosure without any correction applied. A dedicated real-time correction framework for compensating enclosure-induced LiDAR refraction has been presented in prior work, where beam deviations were modelled using analytical ray tracing and physics-based simulation, and corrected using a lookup-table implementation. That approach reduced dimensional error by approximately 40–45% at a distance of 5 m, with negligible computational overhead. In the present work, the focus is placed on characterising the enclosure-induced effects within the integrated hardware system, while the detailed correction methodology is referred to the previous study [27].

In addition to dimensional accuracy, the LiDAR's capability to resolve closely spaced objects was evaluated. A custom-designed target, consisting of three distinct rows of elements with varying separations and orientations, was mounted on a rigid board, as shown in Figure 11(a). Each row was configured to represent different levels of separation and orientation, simulating realistic conditions that the sensor might encounter in the field. The target was placed at a fixed distance of 3 m from the LiDAR, and a scan was performed with the enclosure mounted on the sensor. The resulting point cloud is presented in Figure 11(b).

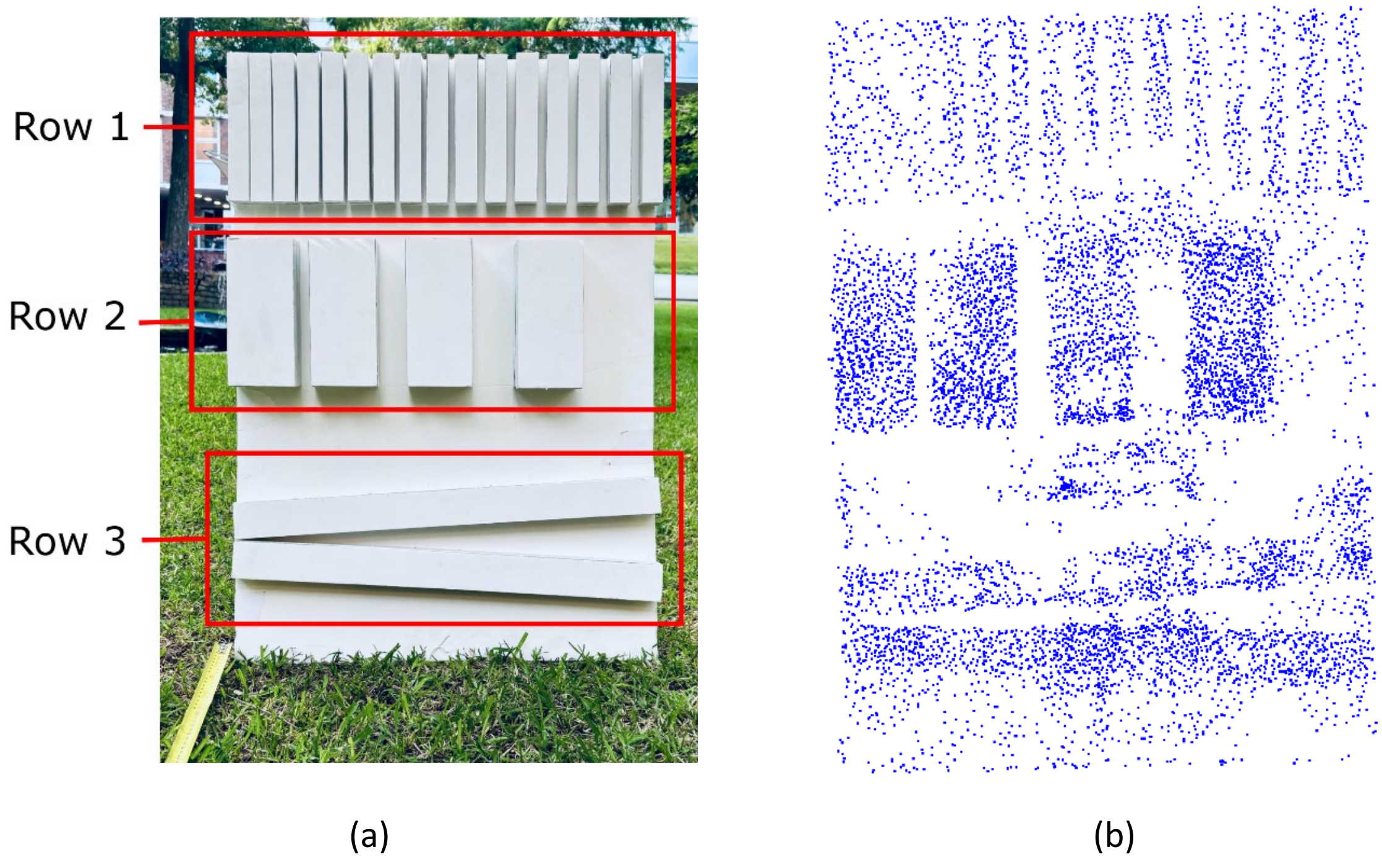

*Figure 11: Small objects separation assessment (a) A custom-designed target featuring three rows to evaluate the LiDAR's resolution capabilities. The top row consists of elements with separations ranging from 1–25 mm, the middle row consists of elements spaced between 30–50 mm, and the bottom row consists of a "<" shaped pattern. (b) The resulting LiDAR point cloud was acquired at a distance of 3 m from the targets, and background points from the supporting board were partially removed for clarity.*

As illustrated in Figure 11, the smallest separations in the top row, particularly those below 10 mm, are not clearly resolved and tend to merge into the point cloud, indicating the sensor's practical resolution limit under the given conditions. However, as the spacing within the same row increases (up to 25 mm), the separations become progressively more distinguishable. The middle row, with separations between 30 mm and 50 mm, exhibits clearly defined, well-separated clusters, demonstrating stable performance under moderate spacing conditions. The bottom row, which consists of angled elements, further demonstrates the LiDAR's ability to capture non-orthogonal geometries. The resulting point cloud preserves the structural orientation of these elements, confirming that the sensor can effectively represent objects with varying orientations.

Overall, while limitations are observed at very small separations, the LiDAR consistently resolves objects with gaps exceeding approximately 30 mm, indicating reliable performance for typical structural features encountered in underground environments.

### 4.2 Camera

To evaluate the optical impact of the enclosure, images were captured under identical conditions both with and without the enclosure in place. Although the two sets appear visually similar, quantitative analysis based on camera calibration reveals systematic and measurable differences across all evaluated metrics, as summarised in Table 2.

*Table 2: Comparison of camera calibration results with and without enclosure.*

| | | Without enclosure | With enclosure |
|---|---|---|---|
| Captured image | | | |
| Colour difference ΔRGB | | 0.0 | 16.61 (+16.61) |
| Sharpness | | 16.19 | 13.03 (-3.16) |
| Edge strength | | 0.015 | 0.014 (−0.001) |
| Line straightness (deg) | | 0.444 | 0.998 (+0.554) |
| Camera Intrinsics | $f_x$ | 231.14 | 224.23 (−6.91) |
| | $f_y$ | 230.66 | 226.22 (−4.44) |
| | $C_x$ | 240.69 | 246.68 (+5.99) |
| | $C_y$ | 182.70 | 178.79 (−3.91) |
| Camera distortion parameters | $K_1$ | −0.246114 | −0.262286 (−0.016172) |
| | $K_2$ | 0.097548 | 0.094721 (−0.002827) |

To quantify the colour difference introduced by the enclosure, ΔRGB is computed as:

$$\Delta RGB = \sqrt{(R_1 - R_2)^2 + (G_1 - G_2)^2 + (B_1 - B_2)^2}$$

where $R$, $G$, and $B$ represent the mean pixel intensities of each colour channel in the two images. For the reference case without the enclosure, ΔRGB = 0.0. With the enclosure, ΔRGB increases to 16.61, indicating a noticeable shift in the overall colour content. This increase occurs because the enclosure modifies the light before it reaches the sensor. In particular, different wavelengths are transmitted unevenly, leading to small but consistent changes across the RGB channels. In addition, reflections within the enclosure further alter the incoming light. As a result, the average channel intensities shift, producing the observed increase in ΔRGB.

Image sharpness is defined as the mean gradient magnitude across all pixels:

$$\text{Sharpness} = \frac{1}{N} \sum_{x,y} \sqrt{(G_x(x,y))^2 + (G_y(x,y))^2}$$

where $G_x$ and $G_y$ are the horizontal and vertical intensity gradients at each pixel, and $N$ is the total number of pixels. Without the enclosure, a sharpness value of 16.19 is obtained. With the enclosure installed, this decreases to 13.03, representing an approximately 19% reduction. This degradation occurs because the enclosure surface introduces scattering and mild blur, which smooth out fine spatial intensity variations and weaken the overall gradient response.
Edge strength follows a similar trend and is defined as the mean gradient magnitude computed over detected edge pixels:

$$\text{Edge Strength} = \frac{1}{\mid E \mid} \sum_{(x,y) \in E} \sqrt{(G_x(x,y))^2 + (G_y(x,y))^2}$$

where $E$ is the set of detected edge pixels. This metric decreases slightly from 0.015 to 0.014 after introducing the enclosure. Although the change is small, it shows that even strong edges in the image become slightly weaker. This means the difference between object boundaries and their surroundings is reduced.

Line straightness measures the weighted angular deviation of detected lines from the principal axes, and is defined as:

$$\text{Straightness} = \frac{\sum_i w_i \cdot \min\left(\mid \theta_i \mid, \mid \theta_i - 90^\circ \mid\right)}{\sum_i w_i}$$

where $\theta_i$ is the angle of the $i$-th detected line and $w_i$ is its corresponding weight. Lower values indicate that the lines are closer to being perfectly horizontal or vertical, meaning less geometric distortion. Without the enclosure, a value of 0.444° is obtained. With the enclosure installed, this increases to 0.998°, more than doubling the deviation. This increase shows that straight lines in the scene are no longer projected as perfectly straight in the image. The main reason is refraction at the curved enclosure surface, which slightly bends the light rays before they reach the sensor, causing the observed geometric distortion.

The intrinsic calibration parameters show consistent changes when the enclosure is introduced. The focal lengths $(f_x, f_y)$ decrease from 231.14 and 230.66 to 224.23 and 226.22, corresponding to reductions of approximately 3.0% and 1.9%, respectively. This reduction corresponds to a slight widening of the effective field of view, consistent with refraction through a curved dome. The principal point $(c_x, c_y)$ shifts by changes of +5.99 pixels in the horizontal direction (right) and −3.91 pixels in the vertical direction (upward). This suggests a mild misalignment between the camera and the geometric centre of the enclosure, resulting in direction-dependent refraction.

The radial distortion coefficients $(k_1, k_2)$ also vary, with $k_1$ changing by roughly 6% relative to its original magnitude and $k_2$ exhibiting a smaller percentage variation of 3%. The increase in $|K_1|$ indicates stronger barrel distortion. This behaviour is consistent with the geometric distortion already reflected in the increase in line straightness from 0.444° to 0.998°, confirming that straight lines are more distorted when the enclosure is present. Together, these results show that the enclosure introduces additional radial bending of light, which is captured by the calibration process. Although these changes are moderate, they are systematic and non-negligible. This confirms that the enclosure alters the effective intrinsic parameters and that calibration must be performed with the dome installed, treating the camera and enclosure as a single integrated optical system.

With the hardware tested and finalised, attention turns to the software that operates on this device. In the next section, we outline the software architecture and capabilities developed with the hardware.

## 5. Software Package

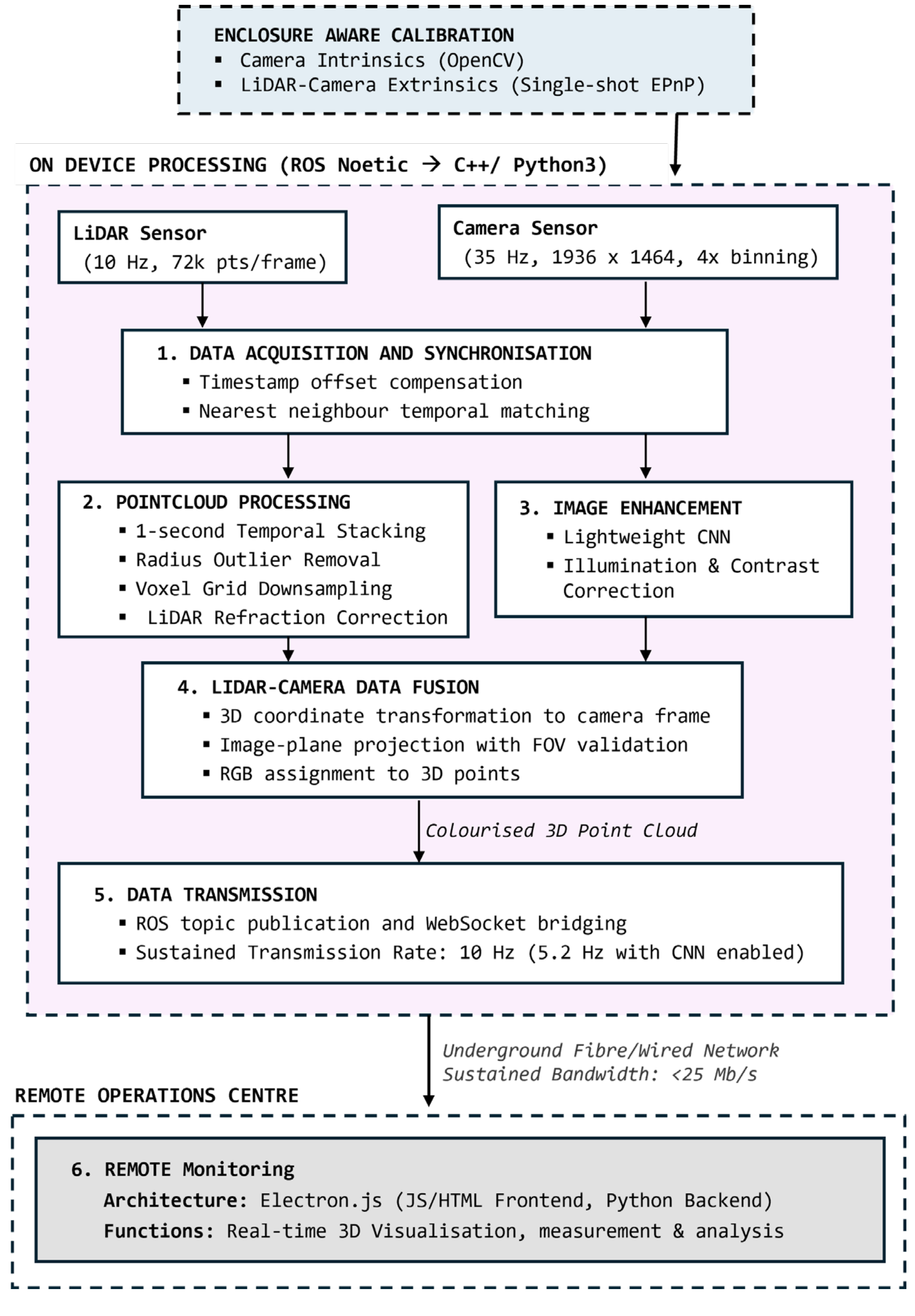

*Figure 12: Software architecture of the LiDAR–camera monitoring system. After enclosure-aware calibration, LiDAR and camera data are time-aligned, processed, and fused onboard to produce colourised 3D point clouds, which are transmitted to the surface workstation for real-time visualisation and analysis.*

The hardware described above is supported by a comprehensive software package developed in the ROS Noetic (Robot Operating System) framework. The core perception pipeline, including data acquisition, synchronisation, preprocessing, filtering, fusion, and transmission modules, is implemented in C++ to ensure deterministic execution and computational efficiency on the onboard single-board computer. The image enhancement module is implemented separately in Python 3 and integrated into the ROS pipeline. Although a complete description of the software architecture is beyond the scope of this paper, its principal components, illustrated in Figure 12, are outlined below.

1. Enclosure Aware Calibration

Protective enclosures introduce optical distortions that affect both LiDAR measurements and camera imaging. As described in our previous work on enclosure-aware calibration, refraction at the air–polycarbonate interface alters ray directions, invalidating standard calibration assumptions. Therefore, the enclosure must be treated as part of the sensing system, and all calibration procedures are performed with the enclosure installed so that these effects are inherently captured [27].

The camera is calibrated using the standard OpenCV checkerboard-based method [28]. A planar checkerboard target of size 42 × 30 cm is used, and multiple images are captured at distances between 1 m and 3 m. These images are used to estimate the intrinsic parameters, including focal lengths, principal point, and lens distortion coefficients, which define the camera's imaging geometry. For LiDAR–camera alignment, the extrinsic transformation is obtained using a single-shot calibration method [29], where the LiDAR point cloud is converted into a 2D intensity image and checkerboard corners are detected in both modalities. Corresponding feature points are then used to compute the rigid-body transformation using the EPnP algorithm.

Calibration quality is evaluated using reprojection error, defined as the pixel distance between projected 3D points and their corresponding 2D image features. For the camera intrinsic calibration, the average reprojection error is approximately 0.08 pixels, indicating accurate estimation of the imaging parameters. For the LiDAR–camera extrinsic calibration, the projection error is approximately 2 pixels, demonstrating good alignment between the two sensors. These calibrated parameters are then used to project LiDAR points onto the image plane and generate colourised 3D point clouds.

2. Data Acquisition and Synchronisation

Data acquisition is implemented in ROS Noetic using two driver nodes operating in parallel, one for the LiDAR and one for the camera. The LiDAR driver aggregates raw packets into full point cloud frames at 10 Hz, producing 72,000 points per frame, each containing $(x, y, z)$ coordinates and intensity values. The camera driver, implemented using the FLIR Spinnaker SDK, captures images at a native resolution of 1936 × 1464 pixels at 35 Hz, below the sensor's maximum rate of 43 fps, to maintain stable GigE throughput through the internal network switch. Pixel binning with a factor of 4 is applied before processing to reduce computational load and bandwidth.

Both sensors assign timestamps at acquisition using their independent internal clocks. As illustrated in Figure 13, the Livox Avia LiDAR continuously accumulates data packets over its acquisition window and assigns a single timestamp at the start of the scan. In contrast, the camera timestamps each frame at the beginning of its exposure interval, which represents a comparatively shorter capture window. Due to these fundamentally different acquisition mechanisms, together with unsynchronised clocks, a systematic temporal offset arises between the two data streams.

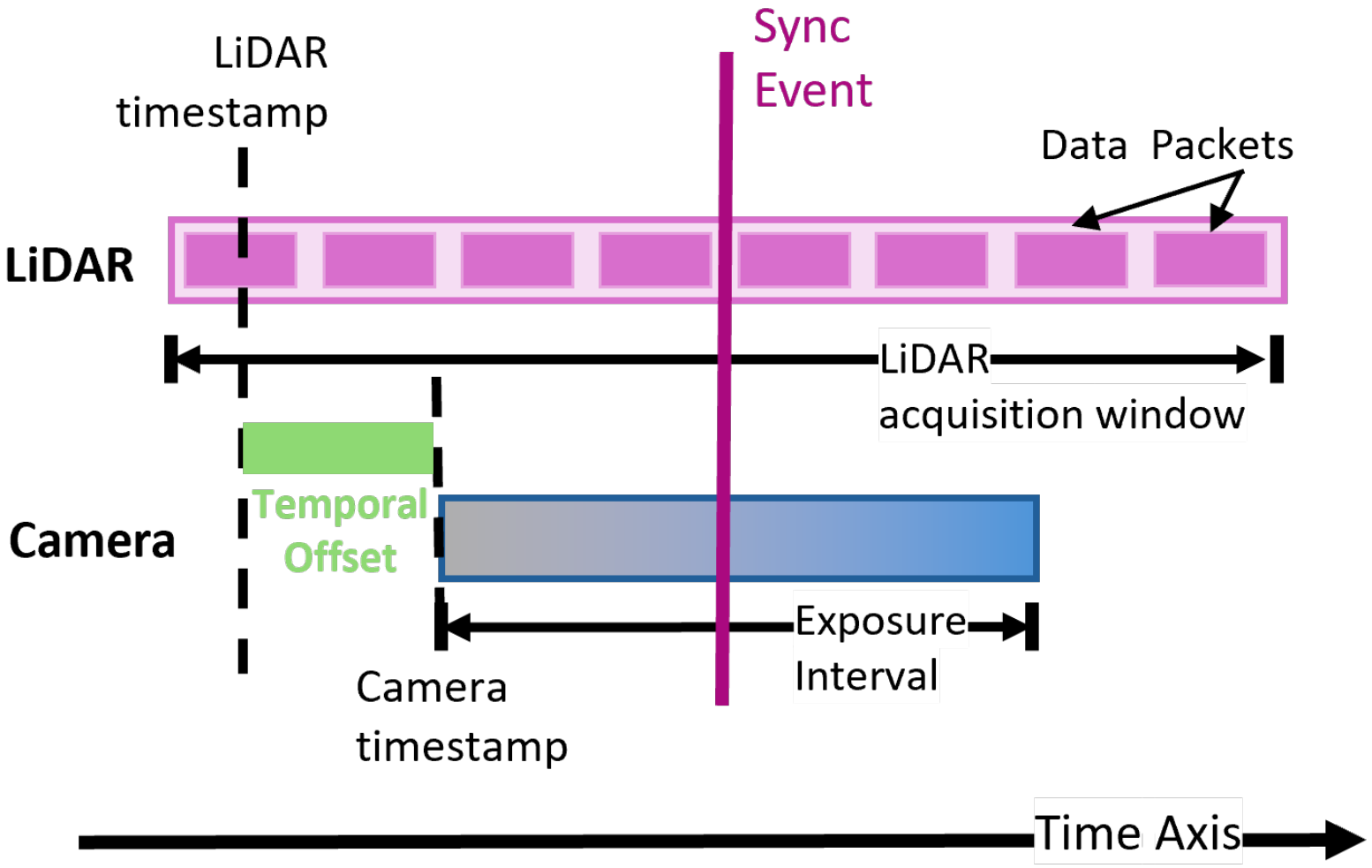

*Figure 13: Temporal relationship between LiDAR and camera data streams. Adapted from [27].*

To address this, the timestamp offset is estimated offline by analysing shared motion events observed in both modalities using our previously published observation-based calibration method [27]. Using this approach, the offset is estimated to be approximately 32 ms, with LiDAR measurements effectively corresponding to an earlier observation time than the associated camera frames. The estimated offset is then used to correct the LiDAR timestamps, aligning them with the camera timeline and reducing the residual synchronisation error to below 3 ms. During runtime, these corrected timestamps are used to associate each LiDAR frame with the temporally closest camera frame using a nearest neighbour matching strategy, ensuring accurate temporal alignment and improved sensor fusion.

3. Point Cloud Processing and Image Enhancement

Following temporal synchronisation, LiDAR frames within a one-second window are accumulated to increase spatial density and improve geometric continuity of the reconstructed scene. The stacked point cloud is first processed using a radius outlier removal filter, in which points with fewer than 5 neighbours within a 0.20 m radius are discarded. A voxel grid filter is then applied to achieve uniform point density and reduce computational load, with a default voxel size of 0.01 m, which is user-configurable depending on the required spatial resolution and bandwidth constraints. Under these conditions, each LiDAR frame contains approximately 72,000 points, and accumulation over a one-second temporal window produces a combined point cloud of approximately 720,000 points. Subsequent voxel-based downsampling reduces this to approximately 100,000–150,000 points per second, maintaining structural detail while ensuring efficient data handling for downstream processing and transmission.

To compensate for enclosure-induced LiDAR refraction, a correction stage is applied to the point cloud using precomputed angular offsets derived from prior calibration. These corrections are implemented through a lookup-table mapping, enabling direct adjustment of LiDAR beam directions prior to data fusion. As observed in Section 4.1, the uncorrected measurements exhibit relative errors of up to approximately 2% (Figure 10). The application of this correction reduces these errors to below 1%, improving the geometric consistency of the reconstructed point cloud and mitigating distortion introduced by the polycarbonate dome [27].

The synchronised camera images are enhanced prior to fusion using a lightweight convolutional neural network integrated within the ROS Noetic pipeline. The model consists of a shallow convolutional architecture with a residual update mechanism that learns to predict an illumination correction map directly from the input image. This predicted map is then used to adaptively adjust pixel intensities, increasing brightness in darker regions while preserving structural details and contrast [24]. The network is highly compact (~40 KB) and runs on the Rock 5A's CPU with an average inference time of ~140 ms per frame, providing improved visibility and contrast in low-light underground conditions while remaining computationally efficient for onboard deployment.

4. LiDAR–Camera Data Fusion

This stage integrates the synchronised LiDAR and camera data to generate colourised 3D point clouds. For each LiDAR frame, the 3D points are first transformed from the LiDAR coordinate frame into the camera coordinate frame using the calibrated extrinsic transformation matrix. The transformed points are then projected onto the image plane using the intrinsic camera parameters.

For each projected point, the corresponding pixel location is computed and validated to ensure it lies within the image bounds and in front of the camera. Valid points are assigned RGB values from the synchronised camera image, while points falling outside the field of view or behind the camera are discarded. The colour information is then appended to the original 3D coordinates, resulting in a geometrically consistent, colourised point cloud in the LiDAR reference frame.

5. Data Transmission

The fused data streams are published as ROS Noetic topics within the onboard system, with colourised point clouds transmitted at up to 10 Hz, matching the LiDAR frame rate under nominal operation. The data are serialised using standard ROS message formats (sensor_msgs/PointCloud2) and streamed in real time to the surface monitoring workstation via ROS Bridge over the mine's wired fibre network using TCP/IP. Bandwidth is managed through voxel-based downsampling and controlled point cloud density, without applying additional lossy compression. To evaluate the effect of voxel size on transmission performance, the voxel grid size was varied from 0.005 m to 0.05 m. As shown in Figure 14, smaller voxel sizes preserve higher point density but increase bandwidth demand. The 0.005 m setting exceeds the 25 Mb/s reference network capacity and reduces the achievable update rate, whereas voxel sizes of 0.01 m and above support stable real-time streaming. The 0.01 m setting provides the highest point density while remaining within the bandwidth limit, making it the default configuration for standard operation.

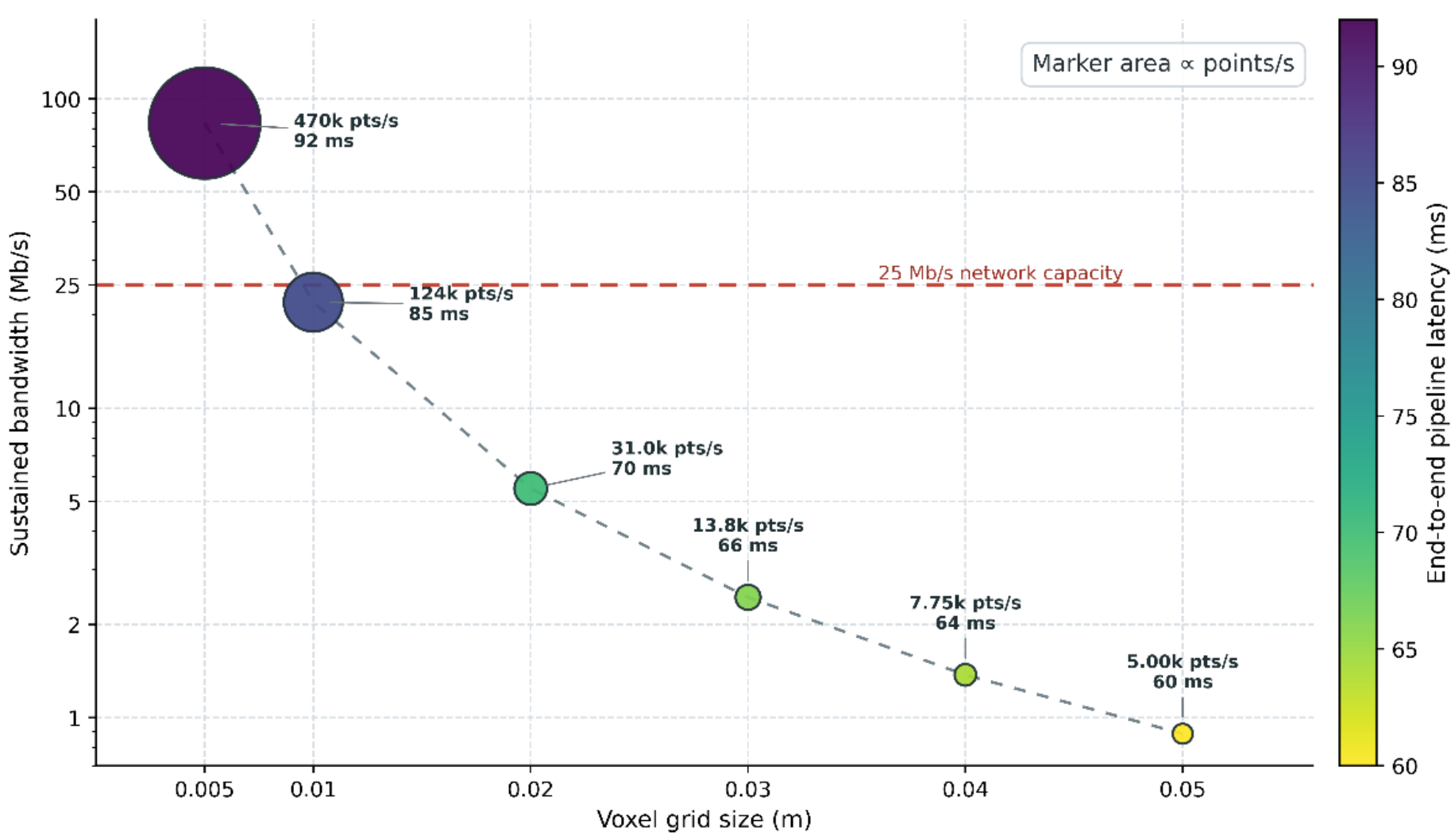

*Figure 14: Effect of voxel size on point density, bandwidth, and end-to-end pipeline latency.*

Under the default 0.01 m voxel setting and without low-light enhancement, the end-to-end pipeline latency was approximately 85 ms, corresponding to about 11.8 frames per second. This provides sufficient processing margin for stable 10 Hz transmission, matching the LiDAR frame rate during nominal operation. When the low-light image enhancement module is enabled, the effective transmission frequency decreases to 5.2 Hz due to additional computational demand. This enhancement module is enabled selectively based on lighting conditions in the mine and is not required under normal illumination. Although the update rate decreases in this mode, 5.2 Hz remains adequate for practical remote monitoring and operator decision support in underground operations [18]. The enhancement network runs entirely on the CPU of the single board computer, and thermal testing confirmed that the additional processing load remains within certified enclosure temperature limits, ensuring stable, continuous operation. Together, these stages complete the onboard processing and transmission pipeline, after which the optimised data stream is received by the surface-level monitoring infrastructure.

6. Remote Monitoring

At the surface level, the incoming ROS Bridge WebSocket streams are received by a dedicated workstation acting as the remote operations centre. The colourised LiDAR point clouds are reconstructed and visualised up to 10 Hz, consistent with the transmission rate. Following onboard optimisation and preprocessing, the sustained bandwidth remained below 25 Mb/s. This is within the capacity of many underground wired communication systems, although available bandwidth remains mine-specific and depends on the deployed network infrastructure [19]. A dedicated surface-side toolkit has been developed to receive, visualise, and manage the incoming data streams. The application enables real-time 3D visualisation, interactive inspection, and measurement within the reconstructed scene, supporting effective remote monitoring of underground operations. It is designed to interface with existing mine monitoring systems, allowing the fused LiDAR–camera data to be integrated into operational workflows and extended for further analysis and automation tasks.

In summary, the software system converts asynchronous raw LiDAR and camera streams into temporally aligned, geometrically consistent, and colourised 3D representations of the underground environment in real time. Through calibrated projection, controlled preprocessing, and structured transmission, the pipeline delivers spatially accurate data that can be directly used for remote assessment and operational decision support. The system enables continuous monitoring of roof geometry, equipment positioning, and material distribution from the surface, reducing reliance on manual inspection and improving situational awareness.

## 6. Experimental Validation

As an intermediate validation step before on-site installation, a controlled simulation campaign was conducted to evaluate the complete sensing and fusion workflow under representative longwall geometry and lighting conditions. This stage enabled the system configuration, sensor placement, low-light imaging, point cloud processing, and end-to-end data transmission to be assessed in a repeatable environment before progression to field deployment. Full-scale testing in an operational underground longwall panel is considered as part of future work, subject to site access, installation approval, and production scheduling.

A simulated underground longwall environment was developed in ROS Gazebo to replicate representative tunnel geometry and operational behaviour. The longwall section measured 5 m in width and 18 m in length, with the roof height set to 3 m to reflect typical underground clearances as illustrated in Figure 15. A shearer model was introduced and animated to traverse along the longwall face, emulating cutting movement and enabling evaluation under dynamic scene conditions.

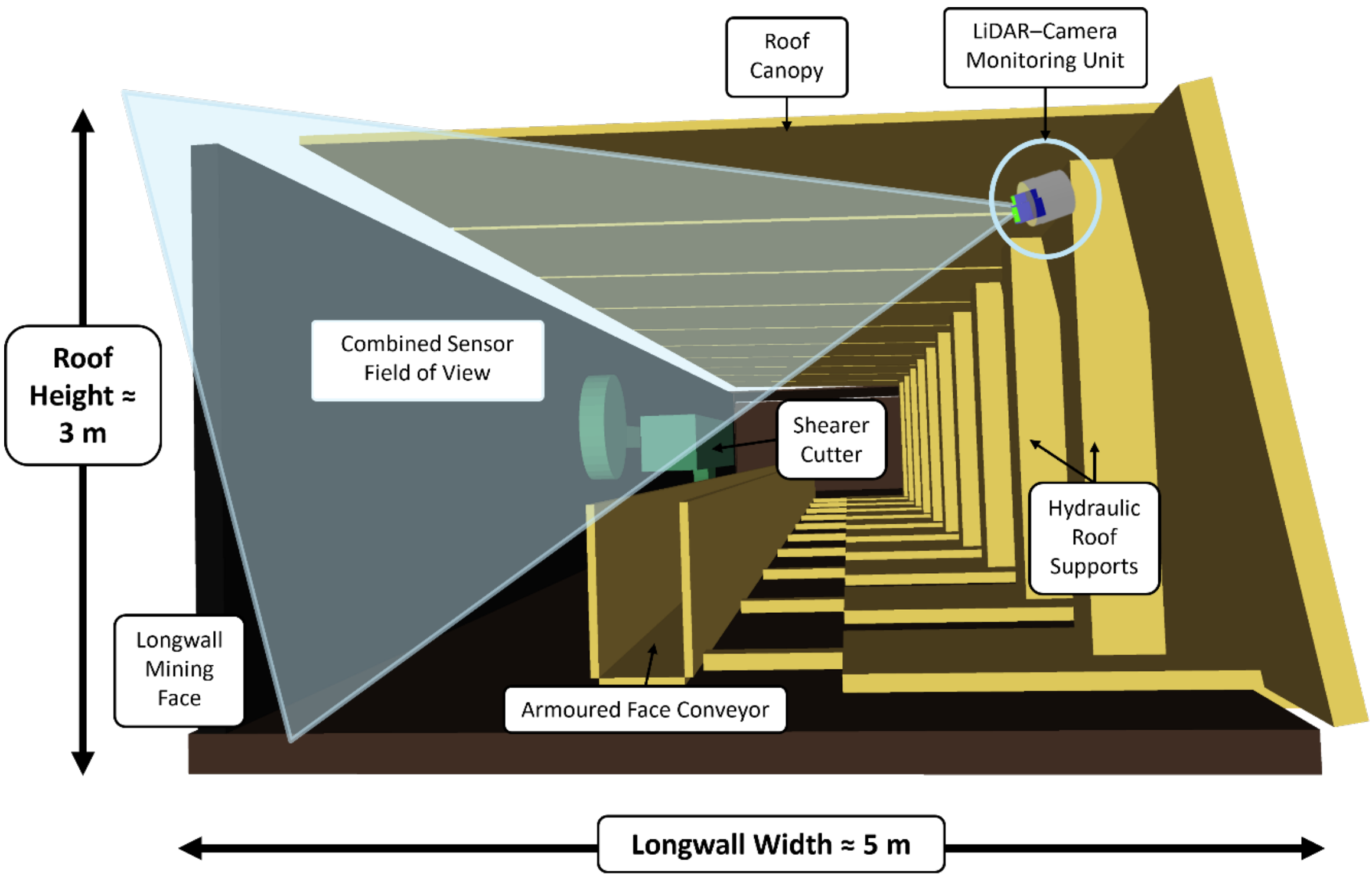

*Figure 15: Simulated longwall environment in ROS Gazebo showing key underground components and the LiDAR–camera monitoring unit mounted beneath the roof canopy, positioned toward the longwall mining face.*

Illumination levels were configured to approximate reduced underground lighting conditions. The simulated environment was adjusted to produce average illuminance levels in the range of 5–10 lux, representative of dim operational conditions in longwall panels. This configuration enabled evaluation of the image enhancement module under low-light conditions [30]. A comparison between the raw camera image and the enhanced output is shown in Figure 16(a) and Figure 16(b). On the 8-bit intensity scale, the mean image brightness increased from approximately 34 to 120, corresponding to an improvement of about 253%. This substantially improved the visibility of structural features before LiDAR–camera fusion.

The sensing unit was modelled using Xacro-based URDF descriptions parameterised through YAML configuration files to match the physical enclosure design. The dome was represented as a visual hemisphere to preserve optical visibility while maintaining geometric realism. The device was positioned beneath the roof supports and oriented toward the mining face, consistent with the intended deployment placement. The camera was mounted inside the dome using calibrated positional and rotational offsets, and was configured with the effective field of view of the physical system, namely 84.8° horizontally and 65.3° vertically (101.6° diagonally), together with a binned resolution of 484 × 366 pixels and a frame rate of 35 Hz. Similarly, the LiDAR was positioned according to the physical assembly configuration and operated at 10 Hz with a sensing range of 0.1–200 m, a horizontal field of view of 70.4°, and a vertical field of view of 77.2°, consistent with the specifications of the deployed sensor. The non-repetitive scanning behaviour of the Livox Avia was reproduced using the livox_laser_simulation Gazebo plugin, configured with a time-indexed scan profile derived from the sensor's native pattern [31].

Simulated sensor data were processed onboard through synchronisation, frame accumulation, filtering, and LiDAR–camera fusion. Consecutive LiDAR frames were temporally stacked to increase spatial coverage and compensate for sparse sampling. Radius outlier removal was applied to eliminate isolated returns, followed by voxel down-sampling to achieve more uniform point density.

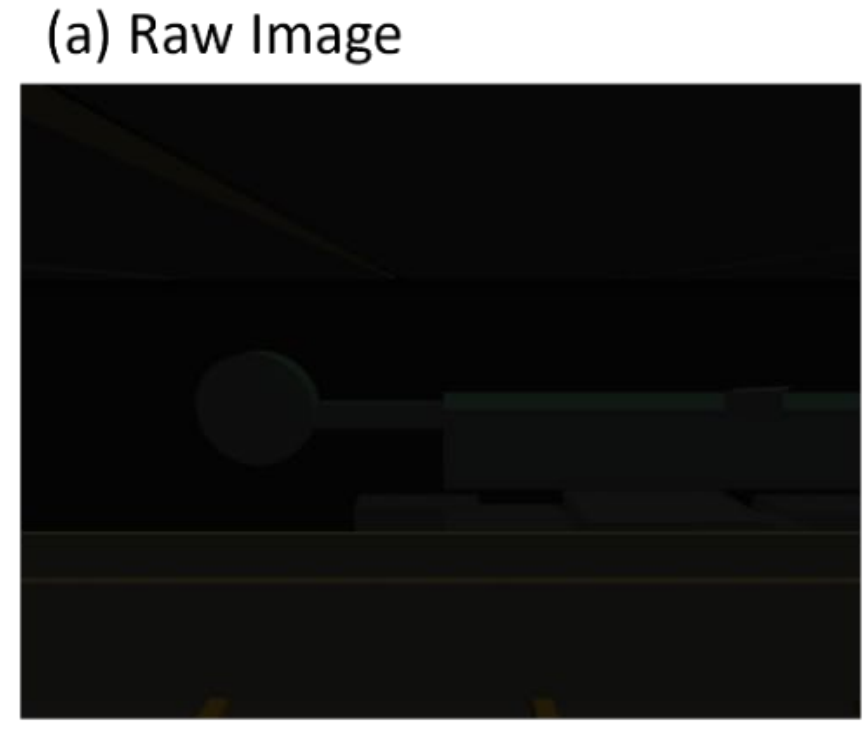

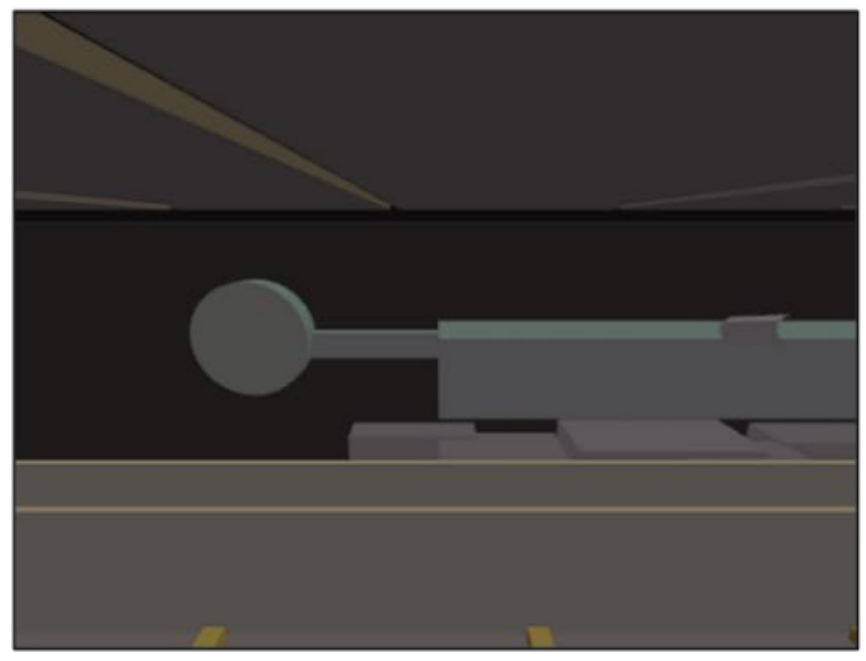

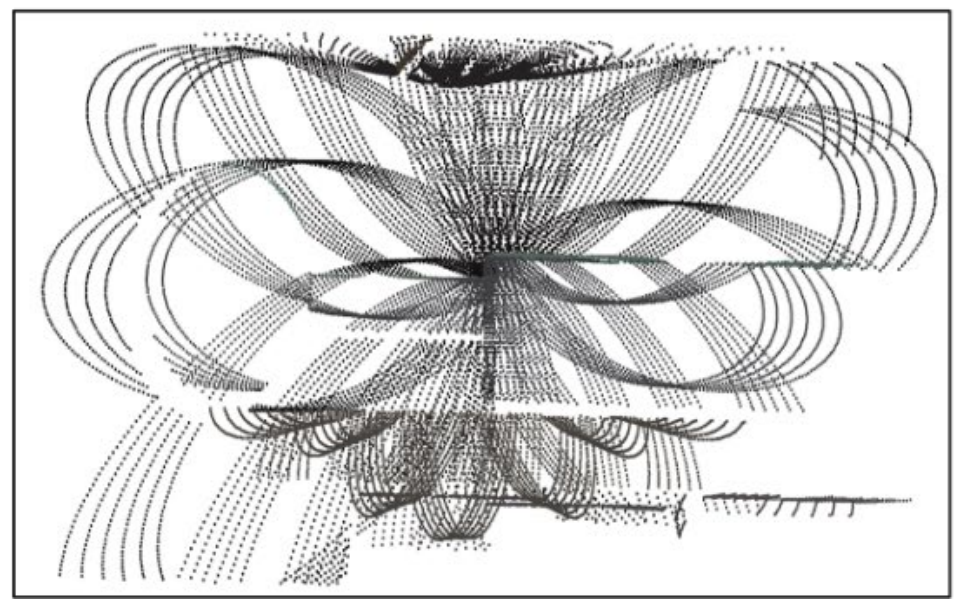

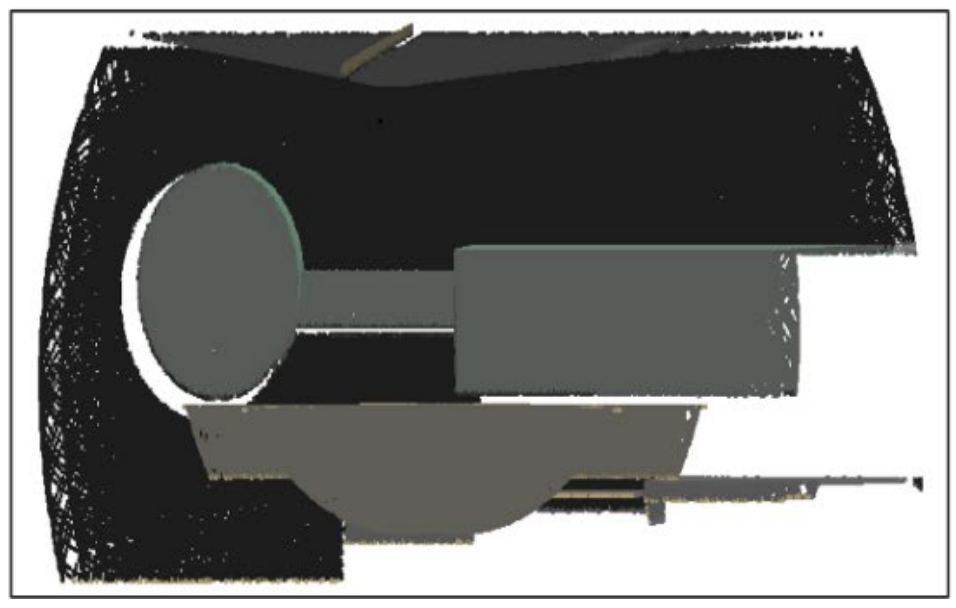

*Figure 16: Image enhancement and LiDAR preprocessing in the simulated longwall environment. (a) Raw low-light image. (b) Enhanced image. (c) Raw colourised point cloud generated from merged sensor inputs. (d) Processed and colourised point cloud after full onboard processing within the enclosure.*

A qualitative comparison of the raw and processed outputs is presented in Figure 16. The raw low-light image exhibits reduced visibility and obscured structural features, while the enhanced image improves contrast and surface definition. The raw LiDAR point cloud shows gaps due to the scanning pattern. After stacking and filtering, the processed point cloud demonstrates improved spatial continuity, reduced noise, and clearer structural boundaries, with point density increasing from approximately 72,000 points per frame to 100,000–150,000 points per second after aggregation and voxel filtering. Integrating the enhanced images with the processed point cloud results in a more coherent and visually consistent colourised 3D reconstruction. The end-to-end processing and transmission pipeline operates with an effective latency of ~80–160 ms, with update rates of 10 Hz without image enhancement and ~6 Hz when the enhancement module is enabled. This closely matches the real system performance, where the update rate is approximately 5.2 Hz with enhancement. The final fused data stream (Figure 16(d)) is then transmitted to a separate workstation for remote monitoring, where the reconstructed scene is visualised in real time, thereby validating both geometric consistency and end-to-end communication stability.

## 7. Discussion

Our design philosophy emphasises selecting commercially available yet robust sensors, paired with a certified flameproof enclosure and intrinsically safe components, to ensure reliable operation even under adverse conditions. Hardware was configured in a compact, multi-layered assembly that optimised the sensors' fields of view and maintained system integrity under high temperatures. Iterative prototyping and controlled thermal testing under representative operating loads enabled refinement of component placement and heat dissipation pathways, ensuring that continuous operation remained within certified temperature limits. Consideration of minimal external cabling and straightforward integration with existing mine infrastructure further supports the system's practicality for sustained underground deployment.

From a software and data-processing perspective, the experimental evaluation confirms that the pipeline achieves a balanced trade-off among spatial fidelity, computational efficiency, and transmission feasibility. Temporal stacking increases structural completeness, while voxel filtering reduces aggregated point density by nearly 80 percent without visibly degrading key geometric features. This reduction is essential for maintaining bandwidth below 25 Mb/s while preserving spatial interpretability. In addition, the enclosure-induced LiDAR distortion identified in Section 4.1 is mitigated by a dedicated correction stage within the processing pipeline, reducing the relative measurement error from approximately 2% to below 1% and improving the geometric consistency of the reconstructed point cloud. Complementing this, the low-light image enhancement module improves visibility in poorly illuminated mining conditions without requiring additional lighting infrastructure. Despite operating entirely on the CPU of the single board computer, the enhancement model maintains stable performance and does not introduce excessive thermal load, demonstrating that intelligent edge processing can be achieved within explosion protection constraints.

In addition to physical experiments on each component in laboratory settings, the sensing principle and reconstruction performance were validated in a custom simulation environment that replicates the intended sensor placement geometry under longwall face conditions. The simulator reproduced the non-repetitive LiDAR scanning behaviour of the Livox Avia and modelled realistic underground surfaces to assess reconstruction density, coverage consistency, and fusion behaviour. This allowed controlled evaluation of how the final colourised reconstructions would appear when the device is mounted within the underground enclosure. The combined laboratory and simulation results provide confidence in the geometric consistency and end-to-end functionality of the system. However, environmental factors such as dust, vibration, humidity, and dome contamination are not fully captured in the simulation and require validation through field deployment.

To further contextualise the capabilities of the proposed system, it is useful to compare it with existing underground scanning solutions. To the authors' knowledge, ExScan is the only commercially available underground 3D scanning system currently used in Australian mining operations, making it a suitable benchmark for comparison. The ExScan employs a mechanically actuated 2D LiDAR to generate 3D visualisations. While effective for periodic mapping, this approach lacks colour information and may introduce spatial discontinuities due to its scan-based acquisition. In contrast, the proposed system integrates a solid-state LiDAR with a high-resolution camera to enable real-time fusion of geometric and colour data, producing dense, colourised 3D point clouds and continuous monitoring without manual scan initiation. A summary comparison between the two systems is provided in Table 3.

*Table 3: Comparison between ExScan and the proposed system*

| Feature | ExScan [6] | Proposed System |
| --- | --- | --- |
| LiDAR type | Mechanical 2D LiDAR (Hokuyo) | Solid-state 3D LiDAR (Livox Avia) |
| 3D reconstruction | Generated via mechanical scanning | Direct 3D acquisition |
| Colour information | Not available | Integrated RGB colourisation |

| Output type | Monochromatic point cloud | Colourised 3D point cloud |
|---|---|---|
| Spatial continuity | May contain gaps due to scanning | Improved continuity via multi-frame fusion |
| Operation mode | Scan-based (manual/periodic) | Continuous real-time monitoring |
| Real-time capability | Limited | Real-time (up to 10 Hz) |
| Suitability for AI tasks | Limited | Supports detection and analysis |

Beyond the comparison in Table 3, the proposed platform also provides a flexible basis for future analytical extensions. The software implementation in C++ and Python supports the integration of custom processing and machine learning modules using widely adopted frameworks such as PyTorch and TensorFlow. The current validation was conducted through laboratory testing and simulation-based evaluation, providing a controlled basis for assessing system performance before underground field testing. Future work will progress the system to controlled testing inside an operational underground mine, where practical factors such as airborne dust, vibration, humidity, dome contamination, installation constraints, and network variability can be assessed. Further development will also investigate multi-unit deployment along the longwall face and automated analysis modules for roof-convergence monitoring, hazard detection, personnel and equipment tracking, and machine-guidance support.

## 8. Conclusion

This paper presented the design, integration, and experimental validation of a LiDAR–camera monitoring system for continuous real-time observation of underground longwall operations within a certified flameproof enclosure. The principal findings are summarised as follows:

(1) Iterative thermal testing at 40 °C reduced the peak surface temperature from 106 °C to approximately 70 °C, while the internal temperature stabilised at 57 °C. The system operated continuously without CPU throttling or performance degradation, confirming stable operation within the rated limits of the selected components.

(2) The polycarbonate dome introduced measurable optical and geometric distortions in both LiDAR and camera measurements. This confirmed that the enclosure must be treated as part of the sensing system, with calibration performed after the dome is installed.

(3) The optimised sensor orientation achieved more than 97% LiDAR–camera field-of-view overlap, supporting reliable colourised point cloud generation. Timestamp calibration reduced the inter-sensor temporal offset to below 3 ms, while the onboard processing pipeline delivered colourised point clouds at up to 10 Hz within typical underground bandwidth limits.

(4) Controlled validation in a representative Gazebo longwall environment confirmed stable end-to-end operation under geometric and low-light conditions. The system maintained real-time colourised 3D reconstruction, with low-light image enhancement supporting improved visual quality in poorly illuminated scenes. These results support progression to underground field validation, where dust, vibration, humidity, and dome contamination will require further assessment.

CRediT authorship contribution statement

**Pasindu Ranasinghe:** Writing – original draft, Data curation, Investigation, Conceptualisation, Methodology, Software, Formal analysis, Visualisation, Validation. **Bikram Banerjee:** Writing – review and editing, Conceptualisation, Validation, Supervision. **Simit Raval:** Writing – review and editing, Conceptualisation, Project administration, Funding acquisition, Resources, Supervision.

Funding Statement

This manuscript is part of the project funded by Azure Mining Technology Pty Ltd., an Australian registered and wholly owned subsidiary of China Coal Technology & Engineering Group (CCTEG) under the Project No 2022-3-KJHZ005.

Declaration of competing interest

This manuscript contains part of TWO provisional patents: A provisional patent filed in Australia titled "Integrated Sensing System for Colourised Three-Dimensional Monitoring Through Enclosure" (Application No. 2026902579) and a separate provisional patent filed in China titled "An enhanced low-light point cloud coloring method and system" (202411295521.4).

Acknowledgments

The authors acknowledge the assistance of Dibyayan Patra (PhD student), Kanchana Gamage (Technical Officer) and Mark Whelan (Senior Technical Officer) from the School of Minerals and Energy Resources Engineering, UNSW, for their valuable support during the experimental development and testing phases of this work.

Declaration of generative AI and AI-assisted technologies in the writing process

During the preparation of this work, the authors used OpenAI's ChatGPT to improve the language and readability of the manuscript. After using this tool, the authors reviewed and edited the content as needed and take full responsibility for the content of the published article.